%Paper: astro-ph/9404013
%From: "Renyue Cen" <cen@astro.Princeton.EDU>
%Date: Thu, 7 Apr 94 15:31:56 EDT

% +--------------------------------------------------------------------+
% |								       |
% |			      TABLES.TEX			       |
% |								       |
% |			Ray F. Cowan  15-Feb-85			       |
% |								       |
% |			  Princeton University			       |
% |								       |
% |			Last Revision: 6-May-85			       |
% |								       |
% |   Macros I find handy for making tables.  See TABLEDOC TEX for     |
% |   a	longer description.  The token-counting	macros are straight    |
% |   from the TeXbook's "Dirty	Tricks"	appendix.		       |
% |								       |
% +--------------------------------------------------------------------+
%
\newbox\hdbox%
\newcount\hdrows%
\newcount\multispancount%
\newcount\ncase%
\newcount\ncols% This is the number of primary text columns in the table.
\newcount\nrows%
\newcount\nspan%
\newcount\ntemp%
\newdimen\hdsize%
\newdimen\newhdsize%
\newdimen\parasize%
\newdimen\thicksize%
\newdimen\thinsize%
\newdimen\tablewidth%
\newif\ifcentertables%
\newif\ifendsize%
\newif\iffirstrow%
\newif\iftableinfo%
\newtoks\dbt%
\newtoks\hdtks%
\newtoks\savetks%
\newtoks\tableLETtokens%
\newtoks\tabletokens%
\newtoks\widthspec%
%
%  Book-keeping	stuff--see how often these macros are called.
%
% 09/17/86 The three lines below were changed as the next three	by H.M.
%\immediate\write15{%
%CP SMSG GJMSINK TEXTABLE -----> TABLE MACROS LOADED, JOB = \jobname%
%}%
\immediate\write15{%
-----> TABLE MACROS LOADED%
}%
%
%  Turn	on table diagnostics.
%
\tableinfotrue%
\catcode`\@=11%	 Allows	use of "@" in macro names, like	PLAIN.TEX does.
%  Debugging aid.  Writes #1 on the
%				     user's terminal and in the	log file.
\def\tstrut{\vrule height16pt depth6pt width0pt}%
\def\|{|}%  Make it easy to get	|'s of type other after	they are later
%	    made active.
\def\tablerule{\noalign{\hrule height\thinsize depth0pt}}%
\thicksize=1.5pt%  Default thickness for fat rules.  The user should feel
%		   free	to change this to his preference.
\thinsize=0.6pt%   Default thickness for thin rules.
\def\thickrule{\noalign{\hrule height\thicksize	depth0pt}}%
\def\ctr#1{\hfil\ #1\hfil}%
%
%
%  Here	are things for controlling the width of	the finished table.
%
\tablewidth=-\maxdimen%
\def\tabskipglue{0pt plus 1fil minus 1fil}%
%
%  Stuff for centering or not.
%
\centertablestrue%
%
%
%
%  \vctr vertically centers its	argument in the	row.
%
\parasize=4in%
\gdef\ARGS{########}%  Produces	the correct number of #'s in the preamble
%		       by the time eveything is	expanded and \halign sees
%		       it.
\gdef\headerARGS{####}%	 Same as \ARGS,	but used in \header macros.
\def\@mpersand{&}%  Allows us to get alignment tab characters later
%		    when we have made the character "&"	an active macro.
{\catcode`\|=13%  Make |'s locally active.
\gdef\letbarzero{\let|0}%  Globally define a macro that	allows us to
%			   keep	active |'s from	being expanded in edef's.
\gdef\letbartab{\def|{&&}}%  This \def will cause active |'s read by
%			     \ruledtable to be converted into double
%			     alignment tabs.
}%  End	of locally active |'s.
{\def\ampskip{&\omit\hfil&}%  This local macro skips a vertical	rule.
\catcode`\&=13%	 Now make &'s into active macros.
\let&0%	 This allows us	to expand \ampskip in the next \xdef without
%	 attempting to expand the & and	getting	an "undefined control
%	 sequence" error.
\xdef\letampskip{\def&{\ampskip}}%  This will cause active &'s read by
%				    \ruledtable	to be converted	into
%				    double tabs	and an \omit'ted \vrule.
}%  End	of locally active &'s.
\def\begintable{%  Here	we make	|'s and	&'s active characters so we can
%		   interpret them as macros.  Note that	this action is
%		   true	only until we encounter	the matching \endgroup
%		   token later at the end of the \ruledtable macro.
   \begingroup%
   \catcode`\|=13\letbartab%
   \catcode`\&=13\letampskip%
   \def\multispan##1{%	We must	redefine \multispan to count the number
%			of primary columns, not	physical columns.
      \omit \mscount##1%
      \multiply\mscount\tw@\advance\mscount\m@ne%
      \loop\ifnum\mscount>\@ne \sp@n\repeat%
   }%  End of \multispan macro.
   \def\|{%
      &\omit\widevline&%
   }%
   \ruledtable%	 Now we	call \ruledtable to do the real	work.
}%  End	of \begintable macro.
\long\def\ruledtable#1\endtable{%
%
%  This	macro reads in the user's data entries
%  and converts	them into a ruled table.
%
%  Important note:  Many macros	and parameters are re-defined here, and
%  these must be kept local to the table macros	to avoid conflict with
%  their use outside of	tables.	 This is done by the \begingroup token
%  macro \begintable and the \endgroup token at	the end	of
%  this	macro.
%
   \offinterlineskip%  Needed to make rules touch each other.
   \tabskip 0pt%  Needed for same reason as \offinterlineskip.
   \def\widevline{\vrule width\thicksize}%  Make outer \vrule's	wider.
   \def\endrow{\@mpersand\omit\hfil\crnorm\@mpersand}%
   \def\crthick{\@mpersand\crnorm\thickrule\@mpersand}%
   \def\crnorule{\@mpersand\crnorm\@mpersand}%
   \let\nr=\crnorule%  A shorter abbreviation.
   \def\endtable{\@mpersand\crnorm\thickrule}%
   \let\crnorm=\cr%  Allows us to use \cr for our own purposes.
%
%  Cause user-typed \cr's to follow a row with a \tablerule.
%
   \edef\cr{\@mpersand\crnorm\tablerule\@mpersand}%
   \the\tableLETtokens%	 Get the user's	extra \let's, if any.
%
%  Put the data	entries	into a token register so we can	scan through them
%  and see what	the user is asking us to do.
%
   \tabletokens={&#1}%	We add an extra	alignment tab to the beginning
%			of the first row to allow for the first	\vrule.
%
%  Now count how many rows are in the table and	return the result in
%  count register \nrows; do the same for columns, and return that
%  in register \ncols.
%
   \countROWS\tabletokens\into\nrows%
   \countCOLS\tabletokens\into\ncols%
%
%  Now do a little arithmetic to convert the number of primary columns
%  into	the number of physical columns that the	alignment preamble must
%  prepare for;	 similarly for rows.
%
   \advance\ncols by -1%
   \divide\ncols by 2%
   \advance\nrows by 1%
%
%  Tell	the user how many rows and columns we found in his data, if he
%  wants to know.
%
   \iftableinfo	%
      \immediate\write16{[Nrows=\the\nrows, Ncols=\the\ncols]}%
   \fi%
%
%  Now we actually go ahead and	produce	the table.
%
   \ifcentertables
      \line{%  The final table comes out as an \hbox of	width the \hsize.
      \hss%  The final table will be centered left-to-right.
   \else %
      \hbox{%
   \fi
      \vbox{%
	 \makePREAMBLE{\the\ncols}%  Generate the preamble.
	 \edef\next{\preamble}%	 This line and the next	line force the
	 \let\preamble=\next%	 expansion of all \ARGS	tokens into the
%				 appropriate number of #'s.
	 \makeTABLE{\preamble}{\tabletokens}%  Go do the \halign here.
      }%  End of \vbox.
      \ifcentertables \hss}\else }\fi%	Finish the centering effect.
%					It is important	that no	spaces
%					follow the two `}' here.
%  }%  End of \line.
   \endgroup%  Return all local	macros and parameters to their outside
%	       values.
   \tablewidth=-\maxdimen%  Reset \tablewidth to normal.
}%  End	of macro \ruledtable.
\def\makeTABLE#1#2{%  Does an \halign for the \ruledtable macro.
   {%  Start of	local parameter	values.
   \let\ifmath0%     These macros would	cause trouble if they were to be
   \let\header0%     expanded in the following \xdef; we \let them be
   \let\multispan0%  equal to a	digit, because digits can't be expanded.
%
%  Set up the width specification here.
%
   \ifdim\tablewidth>-\maxdimen	%
 \widthspec=\expandafter{\expandafter t\expandafter o%
 \the\tablewidth}%
   \else %
      \widthspec={}%
   \fi %
%\out{Widthspec=[\the\widthspec]}%
%\out{Preamble=[\preamble]}%
   \xdef\next{%	 We must force the preamble to be expanded BEFORE the
      \halign\the\widthspec{%
%	 \halign is done;  this	\edef\next{...}\next construction
%		 does the trick.
      #1%  This	is the preamble	text.
      \noalign{\hrule height\thicksize depth0pt}%  Makes the top \hrule.
      \the#2\endtable%	This is	the main body.
%
%     \noalign{\hrule height0.7pt depth0pt}%  Makes the	last \hrule.
      }%  End of \halign.
   }%  End of \next.
   }%  End of local values.
   \next%  This	\next must be outside of the local values, because now
%	   we want those troublesome macros in the \let's above	to have
%	   their normal	actions.
}%  End	of macro \makeTABLE.
\def\makePREAMBLE#1{%  This macro generates the	necessary preamble for a
%		       ruled table with	#1 primary columns.
%		       (Primary	columns	means the number of columns NOT
%			counting those used for	vertical rules.)
   \ncols=#1%  Get the number of columns desired.
   \begingroup%	 Start local parameter definitions.
   \let\ARGS=0%	 This is the key to the	whole thing; it	prevents \ARGS
%		 from being expanded in	the followin \edef's.
   \edef\xtp{\widevline\ARGS\tabskip\tabskipglue%
   &\tstrut\ctr{\ARGS}}%  A 1-column preamble.
   \advance\ncols by -1%  One column has been generated; decrement the
%			  counter.
   \loop%  Append as many further columns as needed to the preamble.
      \ifnum\ncols>0 %
      \advance\ncols by	-1%
      \edef\xtp{\xtp&\vrule width\thinsize\ARGS&\ctr{\ARGS}}%
   \repeat
   \xdef\preamble{\xtp&\widevline\ARGS\tabskip0pt%
   \crnorm}%  Adds the last \vrule.
   \endgroup%  End of local parameters.
}%  End	of macro \makePREAMBLE.
\def\countROWS#1\into#2{%  This	counts the number of rows in #1	by
%			   looking for control sequences that end a row,
%			   e.g., \cr, \crthick,	etc., and puts the result
%			   into	count register #2.
   \let\countREGISTER=#2%
   \countREGISTER=0%
%  \out{In countROWS:  tokens are [\the#1]}%
   \expandafter\ROWcount\the#1\endcount%
}%
\def\ROWcount{%
   \afterassignment\subROWcount\let\next= %
}%
\def\subROWcount{%
%  \out{In subROWcount:	 next is [\meaning\next]}%  Debugging aid.
   \ifx\next\endcount %
      \let\next=\relax%
   \else%
      \ncase=0%
      \ifx\next\cr %
	 \global\advance\countREGISTER by 1%
	 \ncase=0%
      \fi%
      \ifx\next\endrow %
	 \global\advance\countREGISTER by 1%
	 \ncase=0%
      \fi%
      \ifx\next\crthick	%
	 \global\advance\countREGISTER by 1%
	 \ncase=0%
      \fi%
      \ifx\next\crnorule %
	 \global\advance\countREGISTER by 1%
	 \ncase=0%
      \fi%
      \ifx\next\header %
%     \out{In subROWcount:  next=header, ncase set=1}%
	 \ncase=1%
      \fi%
%     \out{In subROWcount:  ncase is [\the\ncase]}%
      \relax%
      \ifcase\ncase %
	 \let\next\ROWcount%
%	 \out{subROWcount---> ncase=\the\ncase}%
      \or %
	 \let\next\argROWskip%
%	 \out{subROWcount---> ncase=\the\ncase}%
      \else %
      \fi%
   \fi%
%  \out{subROWcount--->	NEXT=\meaning\next}%
   \next%
}%  End	of macro \subROWcount.
\def\counthdROWS#1\into#2{%
\dvr{10}%
   \let\countREGISTER=#2%
   \countREGISTER=0%
\dvr{11}%
%  \out{In counthdROWS:	 tokens	are [\the#1]}%
\dvr{13}%
   \expandafter\hdROWcount\the#1\endcount%
\dvr{12}%
}%
\def\hdROWcount{%
   \afterassignment\subhdROWcount\let\next= %
}%
\def\subhdROWcount{%
%\out{In subhdROWcount:	 next is [\meaning\next]}%
   \ifx\next\endcount %
      \let\next=\relax%
   \else%
      \ncase=0%
      \ifx\next\cr %
	 \global\advance\countREGISTER by 1%
	 \ncase=0%
      \fi%
      \ifx\next\endrow %
	 \global\advance\countREGISTER by 1%
	 \ncase=0%
      \fi%
      \ifx\next\crthick	%
	 \global\advance\countREGISTER by 1%
	 \ncase=0%
      \fi%
      \ifx\next\crnorule %
	 \global\advance\countREGISTER by 1%
	 \ncase=0%
      \fi%
      \ifx\next\header %
%\out{In subhdROWcount:	 next=header, ncase set=1}%
	 \ncase=1%
      \fi%
%\out{In subhdROWcount:	 ncase is [\the\ncase]}%
\relax%
      \ifcase\ncase %
	 \let\next\hdROWcount%
%\out{subhdROWcount--->	ncase=\the\ncase}%
      \or%
	 \let\next\arghdROWskip%
%\out{subhdROWcount--->	ncase=\the\ncase}%
      \else %
      \fi%
   \fi%
%\out{subhdROWcount--->	NEXT=\meaning\next}%
   \next%
}%
{\catcode`\|=13\letbartab
\gdef\countCOLS#1\into#2{%
%  \out{In countCOLS:  tokens are [\the#1]}
   \let\countREGISTER=#2%
   \global\countREGISTER=0%
   \global\multispancount=0%
   \global\firstrowtrue
   \expandafter\COLcount\the#1\endcount%
   \global\advance\countREGISTER by 3%
   \global\advance\countREGISTER by -\multispancount
%  \out{countCOLS-->[\the\countREGISTER]}
}%
\gdef\COLcount{%
   \afterassignment\subCOLcount\let\next= %
}%
{\catcode`\&=13%
\gdef\subCOLcount{%
%\out{In subCOLcount: next is [\meaning\next]}
   \ifx\next\endcount %
      \let\next=\relax%
   \else%
      \ncase=0%
      \iffirstrow
	 \ifx\next& %
	    \global\advance\countREGISTER by 2%
	    \ncase=0%
	 \fi%
	 \ifx\next\span	%
	    \global\advance\countREGISTER by 1%
	    \ncase=0%
	 \fi%
	 \ifx\next| %
	    \global\advance\countREGISTER by 2%
	    \ncase=0%
	 \fi
	 \ifx\next\|
	    \global\advance\countREGISTER by 2%
	    \ncase=0%
	 \fi
	 \ifx\next\multispan
	    \ncase=1%
	    \global\advance\multispancount by 1%
	 \fi
	 \ifx\next\header
	    \ncase=2%
	 \fi
	 \ifx\next\cr	    \global\firstrowfalse \fi
	 \ifx\next\endrow   \global\firstrowfalse \fi
	 \ifx\next\crthick  \global\firstrowfalse \fi
	 \ifx\next\crnorule \global\firstrowfalse \fi
      \fi%  End	of \iffirstrow.
\relax%\out{subCOL-->  ncase=[\the\ncase]}
% \out{subCOL-->  next=\meaning\next}
      \ifcase\ncase %
	 \let\next\COLcount%
      \or %
	 \let\next\spancount%
      \or %
	 \let\next\argCOLskip%
      \else %
      \fi %
   \fi%
%  \out{subCOL-->  countREGISTER=[\the\countREGISTER]}
   \next%
}%
\gdef\argROWskip#1{%
%  Deletes the next balanced, undelimited argument from	a
%		  token	list.
% \out{---> Entering argROWskip	<---}
% \out{In argROWskip:  deleted arg is [#1]}%
   \let\next\ROWcount \next%
}%  End	of macro \argskip.
\gdef\arghdROWskip#1{%
%  Deletes the next balanced, undelimited argument from	a
%		  token	list.
% \out{---> Entering arghdROWskip <---}
% \out{In arghdROWskip:	 deleted arg is	[#1]}%
   \let\next\ROWcount \next%
}%  End	of macro \arghdROWskip.
\gdef\argCOLskip#1{%
%  Deletes the next balanced, undelimited argument from	a
%		  token	list.
% \out{---> Entering argCOLskip	<---}
% \out{In argCOLskip:  deleted arg is [#1]}%
   \let\next\COLcount \next%
}%  End	of macro \argskip.
}%  End	of active &'s.
}%  End	of active |'s.
\def\spancount#1{%\out{spancount--->\meaning#1}
   \nspan=#1\multiply\nspan by 2\advance\nspan by -1%
   \global\advance \countREGISTER by \nspan
%  \out{number spancount--->\the\nspan;	\the\countREGISTER}
   \let\next\COLcount \next}%
\def\dvr#1{\relax}%
% \omit\hfil%
% \parindent=0pt\hsize=1.1in\valign{%
% \vfil#\vfil&\vfil#\vfil\cr\hfil\hbox{\ Added to\ }\hfil&%
% \hfil\hbox{\ empty events\ }\hfil\cr}\hfil%
\def\header#1{%
\dvr{1}{\let\cr=\@mpersand%
\hdtks={#1}%
%\out{In header:  hdtks=[\the\hdtks]}%
\counthdROWS\hdtks\into\hdrows%
\advance\hdrows	by 1%
\ifnum\hdrows=0	\hdrows=1 \fi%
%\out{In header:  Nhdrows=[\the\hdrows]}%
\dvr{5}\makehdPREAMBLE{\the\hdrows}%
%\out{In header:  headerpreamble=[\headerpreamble]}%
\dvr{6}\getHDdimen{#1}%
%\out{In header:  hdsize=[\the\hdsize]}%
%\striplastCR{#1}%
{\parindent=0pt\hsize=\hdsize{\let\ifmath0%
\xdef\next{\valign{\headerpreamble #1\crnorm}}}\dvr{7}\next\dvr{8}%
}%
}\dvr{2}}%  End	of macro \header.
\def\makehdPREAMBLE#1{%This macro generates the	necessary preamble for a
\dvr{3}%
%		       ruled table with	\ncols primary columns.
%		       (Primary	columns	means the number of columns NOT
%			counting those used for	vertical rules.
\hdrows=#1%  Get the number of columns desired.
{%  Start local	parameter definitions.
\let\headerARGS=0%
%  This	is the key to the whole	thing; it prevents \ARGS
\let\cr=\crnorm%
%		 from being expanded in	the followin \edef's.
\edef\xtp{\vfil\hfil\hbox{\headerARGS}\hfil\vfil}%
\advance\hdrows	by -1%	One row	has been generated; decrement the
%			  counter.
\loop%	Append as many further rows as needed to the preamble.
\ifnum\hdrows>0%
\advance\hdrows	by -1%
\edef\xtp{\xtp&\vfil\hfil\hbox{\headerARGS}\hfil\vfil}%
\repeat%
\xdef\headerpreamble{\xtp\crcr}%
}%  End	of local parameters.
\dvr{4}}%  End of \makehdPREAMBLE.
\def\getHDdimen#1{%
%\out{In getHDdimen:  Arg 1=[#1]}%
\hdsize=0pt%
\getsize#1\cr\end\cr%
}%  End	of macro getHDdimen.
\def\getsize#1\cr{%
%\out{In getsize:  Arg 1=[#1]}%
%  Here	we have	to check arg#1 and see if the first token in #1	is an
%    \end; if so, we stop, else	we check the width of arg#1.
%  We recall that each arg#1 will be terminated	with a \cr token.
\endsizefalse\savetks={#1}%
%\out{In getsize:  the savetks = [\the\savetks]}%
\expandafter\lookend\the\savetks\cr%
%\out{In getsize:  ifendsize = [\meaning\ifendsize]}%
\relax \ifendsize \let\next\relax \else%
\setbox\hdbox=\hbox{#1}\newhdsize=1.0\wd\hdbox%
\ifdim\newhdsize>\hdsize \hdsize=\newhdsize \fi%
%\out{In getsize:  hdsize=[\the\hdsize]}%
%\out{In getsize:  newhdsize=[\the\newhdsize]}%
\let\next\getsize \fi%
\next%
}%
\def\lookend{\afterassignment\sublookend\let\looknext= }%
\def\sublookend{\relax%
%\out{In sublookend:  looknext = [\looknext]}%
\ifx\looknext\cr %
%\out{In sublooknext:  looknext=cr}%
\let\looknext\relax \else %
%\out{In sublooknext:  looknext/=cr}%
   \relax
   \ifx\looknext\end \global\endsizetrue \fi%
   \let\looknext=\lookend%
    \fi	\looknext%
}%
%
%  Allow the user to make his own names	for crthick, etc.
%
\def\tablelet#1{%
   \tableLETtokens=\expandafter{\the\tableLETtokens #1}%
}%
\catcode`\@=12%	 Change	@'s back to their normal category code.

\def\etal{{\it et al.}\hskip 2.5pt}
\def\cf{{\it cf.}\hskip 2.5pt}
\font\machm=cmsy10 \skewchar\machm='60

\def\refset{\parindent=0pt\hangafter=1\hangindent=1em}
\magnification=1200
\hsize=5.80truein
\hoffset=1.20truecm
\newcount\eqtno
\eqtno = 1
\parskip 3pt plus 1pt minus .5pt
\baselineskip 21.5pt plus .1pt
%
%
%this is a draft: {version number}
%
%\draft{11}
%
%
\centerline{ \  }
\vskip 0.35in
\centerline{\bf HOT GAS IN THE CDM SCENARIO:}
\centerline{\bf X-RAY CLUSTERS FROM A HIGH RESOLUTION NUMERICAL SIMULATION}
\vskip 1.00in
\centerline{Hyesung Kang, Renyue Cen, Jeremiah P. Ostriker and Dongsu Ryu}
\centerline{\it Princeton University Observatory}
\centerline{\it Princeton, NJ 08544 USA}
\vskip 2.0in
\centerline{Submitted to {\it The Astrophysical Journal} on June 25, 1993}
\vskip 0.3in
\centerline{To appear in ApJ, July 1, 1994}
\vskip 0.7in
\vfill\eject

\centerline{\bf ABSTRACT}
A new, three dimensional, shock capturing, hydrodynamic code
is utilized to determine the distribution of hot gas in
a standard CDM model of the universe.
Periodic boundary conditions are assumed:
a box with size $85h^{-1}$Mpc having cell size
$0.31h^{-1}$Mpc, is followed in a simulation with $270^3=10^{7.3}$
cells.
Adopting standard parameters determined
from COBE and light element nucleosynthesis, $\sigma_8=1.05$,
$\Omega_b=0.06$ and assuming $h=0.5$, we find the X-ray emitting clusters,
compute the luminosity function at several wavelengths,
the temperature distribution and estimated sizes as well
as the evolution of these quantities with redshift.
We find that most
($\ge 3/4$) of the total
X-ray ($h\nu>0.3$keV)
emissivity in our box originates in
a relatively small number of identifiable clusters which
occupy approximately $10^{-3}$ of the box volume.
This standard CDM model, normalized to COBE, produces
approximately 5 times too much emission from clusters having
$L_x>10^{43}$erg/s, a not unexpected result.
If all other parameters were unchanged, we would
expect adequate agreement for $\sigma_8=0.6$.
This provides a new and independent argument for
lower small scale power than standard CDM at the
$8h^{-1}$Mpc scale.
The background radiation field
at 1keV due to clusters in this model
is approximately $1/3$ of the observed background which,
after correction for numerical effects,
again indicates approximately 5 times too much
emission and the appropriateness of
$\sigma_8=0.6$.
If we had used the observed ratio of gas to
total mass in clusters, rather than
basing the mean density on light
element nucleosynthesis,
then the computed luminosity
of each cluster would have increased still further,
by a factor of approximately ten.

The number density of clusters
increases to $z\sim 1$,
but the luminosity per typical cluster decreases,
with the result that evolution in the number density
of bright clusters is moderate in this redshift range,
showing a broad peak near $z=0.7$,
and then a rapid decline above redshift $z=3$.
Detailed computations of the luminosity functions in the range
$L_x=10^{40}-10^{44}$erg/s in various energy
bands are presented for both cluster central regions
($r\le 0.5h^{-1}$Mpc)
and total luminosities ($r<1h^{-1}$Mpc) to be
used in comparison with ROSAT and other observational data sets.
The quantitative results found disagree significantly from
those found by other investigators using semi-analytic
techniques.
For example, the total volume emission
from hot cluster gas is found to increase by about
a factor of $1.5$ between $z=0$ and $z=1$, but
for the same CDM model Kaiser (1986) predicted an increase
of a factor of $5.7$, for self-similar evolution of clusters.

We find little dependence of core radius on cluster luminosity
and a dependence of temperature on luminosity
$\log kT_x = A+B\log L_x$,
which is slightly steeper ($B=0.38$)
than indicated
by observations.
Computed temperatures are somewhat higher than obsvered,
as expected, in that COBE normalized CDM has too much
power on the relevant scales.
A modest average temperature gradient is found, with temperatures
dropping to $90\%$ of central values
at $0.4h^{-1}$Mpc and $70\%$ of central values
at $0.9h^{-1}$Mpc.
In these models the decrease of the core radius and temperature
with redshift is significant (in rough accord
with the analytic calculations).
We do not expect to see the same result in open
universe models,
so this property
should provide an important discriminant among cosmological models.

Examining the ratio of gas to total mass
in the clusters (which we find to be anti-biased by
a factor of approximately 0.6),
normalized to $\Omega_b h^2=0.015$,
and
comparing to observations, we conclude, in agreement with S. White,
that the cluster observations argue for an open universe.

\noindent
Cosmology: large-scale structure of Universe
-- hydrodynamics
-- Radiation Mechanisms: Bremsstrahlung
-- X-ray: general
\vfill\eject

\centerline{1. INTRODUCTION}
X-ray emission from clusters
of galaxies provides a powerful
cosmological probe.
When first discovered
(Gursky \etal 1971)
the prodigous luminosities ($\sim 10^{44}$erg/s)
were a cause of some surprise,
but in retropect,
it is clear that we should have expected such objects to exist.
Zwicky (1933) had firmly established the existence
of these assemblies with masses of order
$10^{15}$M$_\odot$ and gravitational radii
of order $1$Mpc.
These numbers, and an assumed gas/total
mass of order $10\%$ lead directly
to the prediction of luminosity
in the $1-10$keV range in excess of $10^{43}$erg/s.

But theoretical considerations alone drive
one to expect that this type of
objects should exist, with the catalogued ``clusters"
being only the extreme and easily identifiable members
of the class.
The growth of perturbations in any variant of the many
theories for the origin of structure leads to the ``breaking"
of nonlinear waves, in a fashion not altogether unlike
the breaking of ocean waves or the
steepening of sound waves into shocks.
Then, allowing for the gas content of the universe,
at the surfaces where these caustics occur (\cf Zeldovich 1970),
real gaseous shocks will form with post shock temperatures
of order $(kT/m_0)=(3/256)(H_0\lambda)^2$.
Intersections of sheets
result in still hotter and denser filaments,
and intersections of
filaments with sheets produce clumps, vertices,
where the density, gravitational potential and temperature are
maximal.
These regions also collect galaxies resulting in optically
identifiable ``clusters of galaxies".
Thus the existence of ``X-ray clusters" could have been anticipated.

As observations of X-ray clusters have improved
[\cf Bahcall (1993) for a recent review],
comparison between the output of theoretical models
for the origin of structure and the rapidly
improving real data has become imperative.
But the problem is computationally difficult.
Given that the core radius
of the typical cluster
is less than $0.5h^{-1}$Mpc and the cluster-cluster
separation of order
$50h^{-1}$Mpc,
a resolution
of $100^3=10^6$ cells is the absolute minimum required
to compute any statistically
valid quantities.
This has been beyond the capacity of our
technology until fairly recently.
Kaiser (1986), in a classic paper on this subject,
says ``We [would] need full three dimensional hydrodynamical
simulations which are not at present a viable proposition."

Thus, other simpler, indirect methods were utilized.
In addition to the first order analytic estimates
of mean quantities (Cole and Kaiser 1989, Cavaliere \etal\ 1991),
another type of approach, which is essentially {\it static}
but can simulate very large volumes has been followed.
To address this problem in an ingeniously economical
if approximate way, Bond \& Myers (1991) modelled all nonlinear
structure using the Press-Schechter formalism
for the growth of dark matter
clusters and then added to each simulated cluster (hot)
gas in virial equilibrium.
This approach
uses idealized clusters, which are laid down in huge volumes according
to the general prescriptions of a cosmological scenario, through mass
functions based on the number density and the distribution of peaks
of the density fluctuations
(Kaiser 1986; Bond 1990; Bond \& Myers 1991; Blanchard \etal\ 1992),
or to the observed cluster abundances and characteristics, which then are
extrapolated backwards in time (Markevitch \etal\ 1991).
Then the projected sky distribution
is derived, and from it the values of observables of interest.
Recent applications of the Press-Schechter formalism
by Evrard \& Henry (1991) and
Blanchard \etal (1992) have
aimed at predicting the ROSAT cluster observations.
Another approach is to adopt
``constrained initial conditions"
(for recent applications of this method see Evrard, Summers \& Davis 1992).
This enables one to use a small volume
which still contains a cluster.
Resolution is gained but statistical information is lost.

We have been following a more ``brute force"
approach with increasingly accurate numerical methods.
In Cen \etal (1990) we used
the Jameson (1989) code with cell number
$N=100^3$ in a box of size $L=30h^{-1}$Mpc to model
the X-ray luminosity function in a CDM scenario.
This was improved on in a
$128^3$ ($L=64h^{-1}$Mpc) simulation using the same code
(Cen \& Ostriker 1992)
and still further with a $N=200^3$, ($L=80h^{-1}$Mpc)
run (Cen \& Ostriker 1993).

In all of these computations the resolution $\Delta l$
was somewhat worse than
$2.5\times (L/N^{1/3})h^{-1}$Mpc, since the
diffusive Jameson code spreads a shock over several cells.
In the three quoted calculations the cell size was approximately
$0.4h^{-1}$Mpc, which thus corresponded to an effective
top-hat resolution of radius
$\Delta r > 0.4\times 2.5\times (3/4\pi)^{1/3}=0.62h^{-1}$Mpc or
a gaussian of $r_g > 0.36h^{-1}$Mpc.
Real clusters often have core radii which are smaller by almost
a factor of two than this scale,
so these calculations suffered significantly from a lack of resolution.
In the earlier papers we attempted to compensate for this
error using extrapolation procedures detailed in Cen (1992).
But of course {\it it is better to simply increase the numerical resolution}.
Recently,
we have developed a shock capturing ``Total Variation Diminishing"
(TVD) code (Ryu \etal 1993),
with considerably improved resolution,
which enables us to do this.
In estimating resolution the ability to
treat a shock (tests shown in Ryu \etal 1993)
is just one factor.
The ability to handle continous flow,
the amount of unwanted numerical viscosity etc
must also be considered.
A detailed description of the new method,
its derivation and origin,
and the tests of its accuracy are reported on elsewhere (Ryu \etal 1993).
It suffices to note here that we expect
that the resolution of the new code
(for a factor four increase in a shock)
is 2-3 cells and superior by a factor 2-3 to our previous work.

Here, we combine this superior code with a larger scale
run, $N=270^{3}=10^{7.3}$, using a slightly
larger box, $L=85h^{-1}$Mpc, and
the new COBE normalization of the CDM spectrum to produce
a reasonably accurate computation of the X-ray emitting properties
of the CDM model.
An estimate of the numerical reliability of our
results may be obtained by comparing the results
of this paper with those of the following paper
(Bryan \etal 1993), which utilizes a totally independent computational
scheme but identical initial conditions and
identical physical modelling assumptions.
The very brightest clusters are intrinsically
rare objects.
This has led us to take larger and larger volumes in
successive simulations in order to
be better able to represent the bright end of the
luminosity function.
We suspect that we have not yet reached a sufficiently large box
in the present run and that our high luminosity
cutoff may be affected by this bias.
Section 2 describes the method and initial
conditions,
\S 3 the results with \S 4 a comparison with observations and
discussions.

\medskip
\centerline{2. METHOD AND INITIAL CONDITIONS}
\medskip
\centerline{\it 2.1 Method}
\medskip
The superiority of the new TVD code to our previous code (\cf Cen 1992)
can be resolved into three aspects.
(1) The TVD code
is able to capture a shock in 2-3 cells rather than 4-7 cells.
The method (Harten 1983) has elements in common with
the Piecewise Parabolic Method (PPM) code
developed by Norman and colleagues
(Colella \& Woodward 1984;
Bryan \etal 1994).
(2) A new variable (entropy $S$) is added to the code,
which enables us to handle very strong (high Mach number) shocks
without generating artificial, non-physical heating of the gas
in regions outside of the shocks.
Such spurious heating is unavoidable in conventional codes,
where only the total energy is used as a variable
and pressure is obtained by
subtracting the kinetic energy
from the total energy.
In most codes
a slight error in the calculation of the total energy and/or
kinetic energy results in
a large error in computing the pressure (or temperature).
(3) A main difference between cosmological applications and
normal laboratory hydrodynamics is the inclusion of the self-gravitational
forces. It is unavoidable that density diffusion
(an inevitable error in all numerical codes)
in a gravitational field does some work, which affects the momentum
and energy calculation. We take special care in handling this
problem such that energy conservation
(represented by the Layzer-Irvine equation)
of the code is not broken and
the inevitable errors are primarily
in the large terms (kinetic and gravitational energy),
not in the relatively smaller thermal energy.

We have performed various tests on the code.
A standard shock tube test shows that the code
resolves the shock in 2-3 cells.
A one dimensional cosmological pancake collapse
calculation indicates that the code conserves
energy at ($1\%$, $0.01\%$) for ($32$, $1024$) cells.
A three dimensional calculation of the standard CDM model
with $128^3$ cells, $L=64h^{-1}$Mpc and $\sigma_8=1$
conserves the energy to $4\%$.
The detailed presentation of the code with various test
results has been given in Ryu \etal (1993).

The simulation reported
on in this paper did not include any atomic processes, i.e.,
no cooling or heating was added, except for the
adiabatic cooling due to the general expansion of the universe,
and ``heating" occurs only due to adiabatic compression
or to entropy generation at shock fronts.
For the hot gas, which we will discuss in this paper,
this approximation is valid since the cooling time exceeds the Hubble time
by a fair margin.
A Courant number of $0.95$ was used, which we found
to be appropriate
for this calculation after various tests.

\medskip
\centerline{\it 2.2 Initial Conditions}
\medskip
We adopt the standard CDM power spectrum
with the transfer function
given by Bardeen \etal (1984).
The following parameters are used:
$n=1$,
$h=0.5$,
$\Omega=1$,
$\Omega_b=0.06$
and
$\sigma_8=1.05$.
Note that the amplitude normalization of
the power spectrum is determined by
COBE observation (\cf Efstathiou, Bond \& White 1992),
parameterized by $\sigma_8$ to translate into conventional notation.
Our box size is $85h^{-1}$Mpc
with $270^3$ cells and $135^3$ dark matter particles so
our nominal resolution is $0.31h^{-1}$Mpc with
our real spatial resolution a factor 2-3 worse than this.
The choice of $\Omega_b$ is
consistent with light element nucleosynthesis
(Walker \etal 1991).

\medskip
\centerline{3. RESULTS}
\medskip
The X-ray clusters in the simulation are identified as follows.
We first calculate the total X-ray luminosity due to thermal Bremsstrahlung:
$$\eqalignno{L_{ff} &= 4\pi \int_\nu \int_V j_{ff} d^3\vec r d\nu \quad \quad
&(\the\eqtno )\cr}$$
\advance\eqtno by 1
for each cell given the cell density and temperature,
where $j_{ff}$ is given by (in units of erg/cm$^3$/sec/hz/sr)
$$\eqalignno{j_{ff}(\nu)=&{1\over 4\pi}{32e^4h\over 3m_e^2c^3}({\pi
h\nu_0(H)\over 3kT})^{1/2}e^{-h\nu /kT}\times \cr
&g_{ff}(T,\nu)[n(H II)+n(He II)+4n(He III)]n(e) \quad , \quad &(\the\eqtno
)\cr}$$
which assumes cosmic
abundances with both hydrogen and helium fully ionized,
and $g_{ff} (T,\nu)$ is the Gaunt factor (Spitzer 1978).
No allowance is made for
line emission, which will be treated in a subsequent paper.

The cells with
the total X-ray luminosity higher than $10^{38}$erg~s$^{-1}$ are
selected as X-ray bright cells. Then we find the local maxima
(by comparing $L_{ff}$ of each X-ray bright cell with that
of 26 neighboring cells)
and identify them as the centers of the X-ray clusters.
Having defined the centers of the X-ray clusters, we go back
to the whole simulation box to define these clusters.
We analyze the simulation in two different ways
(which correspond to spheres of radius $0.5h^{-1}Mpc$
and $1.0h^{-1}$Mpc) as follows.
First, each cluster core
consists $27$ cells (26 cells surrounding the central cell
plus the central cell).
These 27 cells are weighted so that total volume of the cluster equals
the volume of a sphere of radius $0.5 h^{-1} Mpc$
as appropriate for observationally defined X-ray clusters.
That is,
we define
a weight
$w = R_v$ for the central cell and
the 6 cells that share the faces with
the central cell, $w = (3/4)R_v$ for 12 cells that share the
edges with the central cell, and $w = (1/2)R_v$ for 8 cells that shares
only the corners with the central cell, where
$R_v = (4\pi~0.5^3/3)/(20\Delta x^3)=0.839$.
This weighting scheme compensates
for the adoption of slightly too large a volume by lowering
the weight per cell.
For concentrated clusters we may be underestimating
the luminosity by as much as $16\%$
a correction small compared to other errors.
The ``core" luminosities calculated here are from
regions $r~h<0.5$Mpc but the true
X-ray cores as defined by equation (3) are about a factor
of two smaller.
Second, the total X-ray cluster volume
consists $125$ cells (124 cells surrounding the central cell
plus the central cell) and each cell is weighted equally which
gives a volume equivalent to a sphere of radius $0.98h^{-1}$Mpc.

Our ($85h^{-1}$Mpc)$^3$ box at $z=0$ contained
(0,0) clusters with (total, core) luminosity brighter than $10^{45}$erg/s,
(8,1) brighter than $10^{44}$erg/s,
(29,17) brighter than $10^{43}$erg/s,
(115,68) brighter than $10^{42}$erg/s
and (367,264) brighter than $10^{41}$erg/s.
We attribute the lack of clusters brighter than $10^{45}$erg/s
as simply due to the size of our box.
An additional factor of at least two in scale (and 8 in computer resources)
would be required to significantly improve
on the quoted results.

It is convenient to fit the luminosity
function to the three parameter Schechter function
$$\eqalignno{n(L)dL &= n_0 (L/L_*)^{-\alpha} e^{-L/L^*} d(L/L^*)\quad .\quad
&(\the\eqtno )\cr}$$
\advance\eqtno by 1
The raw results for the luminosity functions are shown
in Figures (1a,2a,3a,4a) for total luminosity,
$0.3-3.5$keV, and $0.5-4.5$keV and $2-10$keV bins emitted from
central $0.5h^{-1}$Mpc regions (core) of each cluster,
and Figure (1b,2b,3b,4b) for $1h^{-1}$Mpc spheres.
The figures show the range of cluster properties which
are most accessible to observation (in the range we are able to compute)
$10^{40}$erg/s$\le L_x\le 10^{44}$erg/s and
$0\le z\le 1$.
We have computed approximate Schechter function
[$n(L)dL\equiv n_0(z)(L/L^*)^{-\alpha} exp(-L/L^*)dL$]
fits to the results
with the numerical parameters ($n_0,L^*,\alpha$)
as a function of redshift collected in Table (1)
and the data extended to $z=5$.
Luminosities and temperatures are presented as they would
be seen by observers
near the clusters,
but densities and emissivities are in comoving coodinates.
Also in Table (1) we integrated over the cluster luminosity function,
using the Schechter fit $j_{cl}\equiv n_0L^*\Gamma(2-\alpha)$
(in units of $10^{40}$erg/s/h$^{-3}$comoving Mpc$^3$),
showing the result in the second to last column and give
also in the last column
the total emissivity from the box as $j_{gas}$,
which includes the emission from lower density regions further from cluster
cores than $0.5h^{-1}$Mpc and also
from clusters whose central emissivity
is less than our cutoff value.

We see that the clusters cores (upper panels of Table 1),
as we have defined them,
contained between $1/2$ and $1/4$ of the total X-ray emission
in the regions studied.
The total cluster luminosity (lower panels)
in the box is typically $3/4$ of the X-ray emission
from the box,
although the bright clusters
($L\ge 10^{43}$erg/s) occupy a volume which is less
than $10^{-3}$ of the total,
i.e., the emissivity is very concentrated at bright peaks.
For the total luminosity,
the Schechter $\alpha$ parameter is approximately
$1.5$ with little evolution,
and $\alpha$ for the few keV
bands is typically slightly flatter at $1.4$.
For $\alpha < 2$, as we have noted,
most of the luminosity arises from bright clusters.

We defer a detailed comparison with observations to a subsequent paper.
For the present we compare in Figure (1b)
the bolometric luminosity distribution with that given
by Henry and Arnaud (1991) and in Figure 4b,
the 2-10keV luminosity function of low redshift clusters, as
summarized by Henry (1992, Fig. 1) with our
computational results.
The observational error is approximately a factor of two but the difference
between computed and observed luminosity functions in Figure (4b)
is approximately a factor of 6 (at $L_x=10^{43.5}erg/s$) and
is probably
significant, especially as our numerical procedures tend to
systematically underestimate the predicted luminosity.
Taking this underestimate to be a factor of $1.8$ (\cf Cen 1992 for
procedure),
the discrepancy corresponds (taking $j_\nu\propto \sigma_8^{4.3}$)
to a factor of $1.7$ in amplitude.
Another way to check the consistency of
this scaling scheme is as follows.
A model with $\sigma_8=0.6-0.7$ corresponds approximately to
$z=0.5$ for $\sigma_8=1.05$ model.
Since $n_e$ goes as $(1+z)^3$ and temperature as $(1+z)^{-1}$,
the total luminosity of each cluster will go like
$n_e^2 T^{1/2} \propto (1+z)^{11/2} = 9.3$ (if the physical size
of a cluster is fixed)
or
$n_e^2 T^{1/2} \propto (1+z)^{5/2} = 2.8$ (if the comoving size
of a cluster is fixed).
Since the cluster size is neither fixed in comoving size nor
in physical size, the truth is probably intermediate of these two numbers.
The average of these two numbers is about 6, which is
consistent with our preceding extrapolation.
Alternatively phrased, a model with $\sigma_8\approx 0.6-0.7$
and the same values of
$(H_0,\Omega_b)$ would have produced approximately the correct
X-ray luminosity function.

This result, that standard CDM, as normalized to COBE on large
scales, has too much
power on small scales by $1.5-2.0$, is by now well known
(\cf Davis \etal 1992).
The results of this paper
confirm the common knowledge using a new technique and a new
observational measure.
This conclusion would be greatly strengthened if we were to use
the observed
ratio of gas to total mass in the X-ray emitting regions rather
than to rely, as we have, on light element
nucleosynthesis for the ratio
$\rho_b/\rho_{tot}$.
Following White (1991) we would have
(for $h=0.5$) taken $(\rho_{gas}/\rho_{tot})=0.13$ rather
than the value we derived $0.04$.
Using the larger, observed value our luminosities
would have been larger by a factor of about ten.

The value of $\alpha$ found in our work for the Schechter parameter of
approximately 1.4-1.5,
is slightly smaller than the best fit observational values
1.9-2.0 quoted by Henry (1992).
This difference, as can be noted visually in our
Fig. 4b, is probably not significant.
The values of $\alpha$ quoted in Table 1 are
primarily determined by lower luminosities
than those used in the observational fits.
In overlapping luminosity ranges the computed
and observed values of $\alpha$ are in adequate agreement.

The number density of clusters
peaks at intermediate redshift
and the typical luminosity is (for small redshift)
relatively constant, so there is
a peak emissivity at approximately
$z=0.8$ for the several keV bands.
Thus, crudely speaking,
in this model one expects ``positive" evolution until nearly
$z=1$ and then
negative evolution thereafter.
But the situation is more complex.

Let us first look at the last two columns of Table (1a),
which show the evolution of the total bremsstrahlung emissivity
per unit comoving volume and the integrated cluster emissivity.
The tabulated evolution for these
two measures is roughly parallel,
indicating that the brightest clusters emit a roughly constant fraction
of the total bremsstrahlung radiation.
The peak, seen at redshift $z=0.5-1.0$,
is at first surprising.
The Press-Schechter treatment indicates that
several effects are at work.
The fraction of mass in deep
potential wells can only increase with time and once gas is virialized,
its density (metric, not comoving) does not change,
but merging of clusters will reduce the typical density
without affecting the temperature.

We believe that the peak in emission
at moderate redshift seen is real.
Clusters form as a result of a hydrodynamic shock,
go through a phase of maximum density,
and then reexpand somewhat to reach an
equilibrium system
with somewhat lower density.
The approximate Press-Schechter formalism, which does not
allow for this overshooting, does not
include this effect.
In addition, the peak kinetic energy density occurs,
for the CDM power spectrum, for waves of lengths $\sim 30 h^{-1}$Mpc,
which became nonlinear at $z\sim 0.5$ with our chosen normalization.
In short, it is not quantitatively accurate
to treat the clusters as derived
from a power law form of $P_k$ and as in exact virial equilibrium.
Both ram pressure confinement and subsequent
expansion are also important phenomena,
as well as the merging phenomena treated by the P-S formalism.

The self-similar evolution adopted in the semi-analytic
treatment of Kaiser (1986, see also Blanchard \etal 1992)
predicts an emissivity scaling as
$(1+z)^{5/2}$ for moderate redshifts.
This is a far stronger evolution than
shown in Table 1a (bottom panel)
columns 8 and 9.
These show a positive
evolution to $z=1$ by a factor
of about $1.6$,
whereas the Kaiser prediction is a factor of 5.7.
Thus, it seems that
detailed numerical simulations, such as the present work,
will be required to permit meaningful
comparisons between the observed
and theoretically predicted evolution
of the cluster X-ray luminosity function.
A conclusive test of the argument presented here could be
made if we were to repeat the simulation with an assumed power
law form for $P_k$.
Then self-similar results should be obtained.
This test is in progress.

While the integrated X-ray emissivity evolves fairly slowly over the period
surveyed in Table (1a),
both $L_*$ and $n_0$ evolve more rapidly and in opposite directions,
with more and more lower and lower luminosity
clusters at higher redshifts.
To highlight the negative evolution of the
bright end
of the luminosity function we computed the fifth and sixth columns
of Tables 1b-1d (sixth and seventh columns of Table 1a),
the comoving density of clusters having luminosity greater than
$10^{43}$erg/s and
$10^{44}$erg/s.
For reasons stated earlier (based on our limited box size),
we use the Schechter fit rather than
direct counts to compute these columns.
Comparing columns 4 and 6 (5 and 7 of Table 1a)
we see that, although the total number
density $n_0$
of clusters increases with redshift
(until $z=3$),
the number density
of high luminosity clusters decreases after $z=0.5$.
This is presumably one of
the effects leading to the observational appearance of ``negative evolution"
(i.e., less bright clusters at higher redshifts).
Statistical fluctuations in our results are
still quite significant due to
the limited box size.

Redshift effects strongly exaggerate this
tendency to observe
negative evolution, since higher redshift clusters
tend to have lower temperatures (\cf column 4
of Table 1a and Figures 5, 12) and both
effects can shift the observed ($z=0$)
spectrum out of the high energy X-ray bands.
Note that the negative evolution in the density
of clusters with $L>10^{44}$erg/s is more
and more steep in Tables (a$\rightarrow$ d)
as one looks at higher energy bands.

The emission weighted temperature,
$T_x$ of each cluster is
calculated and the distributions
are shown in Figure (5).
The arrows in each panel indicate the
luminosity weighted average
at the given epoch.
We see that at all epochs the coolest clusters
dominate the statistics (the turnover
at low $T_x$ is presumably caused by our
definition of minimum cell luminosity to
constitute an X-ray bright cell),
but the mean is determined by the high mass, high luminosity
high temperature end of the distribution.
The increase of the mean with increasing
time is for the usual reason, $T_x \propto (H\lambda)^2$,
with the wavelength of nonlinear waves increasing with
time faster than $H$ decreases.
The mean temperatures, indicated by arrows are included
in column 3 of Table (1a).
Looking ahead to Figure (7a) we see the strong correlation
found between $T_x$ and total luminosity
(clusters are shown at $z=0$).

Now let us turn again to the total
cluster luminosity (vs core luminosity)
as shown in the lower panels of the tables and
in figures 1b-7b,
the quantity normally measured by
satellite observations.
The decline in numbers of bright clusters
with redshift is steeper for the total
than for the cores,
because the high redshift clusters are physically smaller.

We can also roughly estimate the effective radii of the clusters
by assuming that the emission has a profile
$$\eqalignno{j &= {j_0\over [1+(r/r_x)^2]^2}\quad \quad &(\the\eqtno )\cr}$$
\advance\eqtno by 1
and determining, from the ratio of the luminosity
of the central cell to the total cluster luminosity,
the value of $r_x$ which would produce this ratio.
We show in Figure (6) the radii
determined in this fashion.
The peaks seen in the panels
of Figure (6) are, of course, artificially
induced by our nominal
resolution limit of $0.31h^{-1}$Mpc,
but the distribution to larger radii should be reasonably accurate.
Arrows indicate the luminosity weighted average values.
Since brighter clusters tend to be resolved,
these numbers should be reliable.
We see a weak trend of increasing size with increasing
time, which is in the theoretically anticipated direction.
Longer wavelengths became nonlinear later
producing larger clusters, and smaller clusters
merge to produce larger clusters with increasing time.

Now in Figures (7a,b)
we show the scatter plots of ($T_X,r_x$) vs $L_{tot}$.
We see that there is a clear correlation between $L$ and $T_x$.
The best straight line fit shown as dashed corresponds to slopes
of $(0.39,0.37)$ respectively.
The solid line indicates the observation (Henry 1992).
But we do not see any correlation
between $L$ and $r_x$.
These two correlations
imply (since $L_x\propto T_x^{1/2} r_x^3\rho^2$)
that $\rho$ is an increasing function of $J_\nu$
($\rho \sim T_x$).
This is expected since higher density peaks contain more energy
and therefore produce stronger shocks.
The predicted temperatures are higher than those
observed and the difference is probably significant
and in the expected sense.
Just as COBE normalized CDM predicts too high a galaxy velocity
dispersion, so it
predicts too high cluster X-ray temperatures.

Finally, we address temperature variations within clusters.
Given our limited resolution there is little that can be
accurately determined from our simulations, but we are able
to compare the central cell (Volume$=3.1\times 10^{-2}h^{-1}$Mpc$^3$)
with the surrounding cells ($3^3-1^3$)
(volume of size $5.9\times 10^{-1}h^{-3}$Mpc$^3$)
and the cells surrounding these cells ($5^3-3^3$, vol$=3.0 h^{-3}$Mpc$^3$).
We define the ratio of the inner cell to
the next cube
as $T_c/T_{shell}$ and show the scatter
diagram in Figure (8).
No trend is seen with luminosity and
the median value, indicated by the dashed line, is $1.3-1.4$.
The cluster gas deviates significantly from
isothermality with a $30\%$ temperature
decline typically found by a radius of
$0.4-0.5h^{-1}$Mpc.
In Figure (9) we compare the temperatures found
in the three regions noted above,
and normalized to the temperature in the central cell.
The large dispersion is indicated by the error bars.

In the development of a one-dimensional Zel'dovich pancake
(Shapiro \& Struck-Marcell 1984; Ryu \etal 1993)
there is a temperature {\it minimum} at the center
of a pancake and we might expect,
on theoretical grounds,
that a similar effect should occur
in real three-dimensional clusters.
(The reason is that the gas, which is shocked first, is put
on a low adiabat and should maintain a lower temperature
and higher density than
surrounding gaseous elements with which
it is in pressure equilibrium.)
In the following paper (Bryan \etal 1993) we can see
pictorial evidence for such shallow minima.
The shallow temperature gradient of the
opposite sign seen in Figures 8 and 9 is,
we believe not inconsistent with
these theoretical and pictorial expectations,
due to the crude averaging used in producing the above figures.

Let us turn away from the consideration of specifically
identified clusters  to ask a more general question concerning
the hot gas.
Is it fairly represented in dense regions or is it ``biased"
-- over represented -- or anti-biased?
This is a question with great cosmological significance.
If we know the ratio of gas (+ galaxies) to total matter
in the clusters by direct observations,
and we know, from light element nucleosynthesis the global
baryon density,
then we can divide the second number by the first to obtain the global
matter density and compare with the
cherished critical density.
This line of argument has been carefully
examined recently by
White (1991) and also reanalyzed by
Babul \& Katz (1993).

The argument depends on knowing whether or not
$\rho_b/\rho_{tot}$ varies significantly from place to place and,
in particular, will this quantity
be found at near its average value in the high density regions
where it can most easily be measured.

Our possibly counter-intuitive results are shown in Figure (10),
where we plot the ratio $(\rho_{IGM}/\rho_{tot})$
vs $\rho_{tot}/<\rho_{tot}>$
smoothed by a gaussian of radius $1h^{-1}$Mpc.
At any value for
the total density there is a wide range of possible
values of $\rho_{IGM}$, but
the high density regions actually
have a {\it lower} than average ratio
of baryons to total mass, whereas the void
regions are gas rich.
In fact in the regions (not shown) which have
$(\rho_{tot}/\langle\rho_{tot}\rangle)$\quad $<0.1$
(more than 90\% underdense), which fill more than half of the box volume
(filling fraction 51\%) the gas
is relatively overdense by
approximately a factor of five
and constitutes 31\% of the total mass
as compared to the global average of 6\%.
The cause of this segregation is primarily
due to gas pressure but in part due to
statistical effects:
in low density regions there will by chance be regions containing
no dark matter particles.
The trend shown however is real.
After a collapse towards a shock/caustic the difference between
the high pressure inner region and low pressure
low density regions outside the caustic causes the gas to
expand out at the speed of sound in an effort to achieve
pressure equilibrium
(recalling also that the universe as a whole is expanding).
This effect, seen earlier in the one-dimensional calculations
of Shapiro and Struck-Marcell (1985),
is also a partial cause of the peak in the cluster
emissivity at $z\approx 1$.
They reach maximum compression at that time and expand
subsequently with consequent
lowering of X-ray luminosity.

Thus, we believe that the phenomenon discovered
in this high resolution CDM simulation will be found to be generic.
The X-ray clusters have a lower than average ratio
of baryons to total mass.
Evrard (1990) and
all our previous work (with other codes and other scenarios)
has shown the same result.
If input from supernovae in cluster galaxies is important
(\cf Yahil \& Ostriker 1973), as may
be implied by the high metal abundance,
then this statement is strengthened.
The additional energy input from supernova winds
can only further contribute to driving hot gas out
of the clusters.

Even without this effect, we see that in the high density clusters,
where $\rho_{tot}/<\rho_{tot}>$ approaches $10^3$,
the gas is under represented by a factor of $1.7$.
Thus, White's (1991) conclusion that $\Omega$
is significantly less than unity is strengthened
and the estimated value of $\Omega$ further is reduced by this factor.

In our discussion of the bias or antibias of the gas,
so far we have weighted all volumes  equally.
It would be more appropriate to look
specifically
at those regions
within $hr<1$Mpc at the centers
of our X-ray clusters.
Figure (11) shows just that.
The histogram under the thin solid line shows
the distribution of
$(\rho_{gas}/\rho_{tot})$
in our tabulated clusters with the median value of this
quantity, $0.040$, indicating by the vertical thin solid line.
But, since the observed sample is (approximately) picked on a
luminosity weighted basis we constructed the histogram
of
$(\rho_{gas}/\rho_{tot})$
using that weighting and display it as the dotted region
with median indicated by the dotted vertical thin line.
The global (assumed) average for
$(\Omega_b/\Omega_{tot})$ is shown
as the heavy vertical line.
We see an antibias of $0.67$ from the
ratio of median to global value
which is reduced to $0.75$ if
we luminosity weight the sample.
Thus the effect mentioned, that clusters have
a lower than average ratio of gas to total
matter, withstands this more careful scrutiny.
The heavy solid histogram indicating the observational
situation is adapted
from Jones \& Forman (1992).
We see that the observed ratio
is higher than the computed ratio
by a factor of 2-3.

\medskip
\centerline{4. EVOLUTION OF THE X-RAY CLUSTERS:}
\centerline{AN OBSERVATIONAL TEST FOR THE CDM SCENARIO?}
\medskip
In Figures (1-4) we showed
the evolution of the cluster
luminosity
function expected in the CDM scenario.
In the easily observed range of parameters
($0\le z\le 1$, $10^{40}$erg/s$\le L_x\le 10^{44}$erg/s)
little evolution is seen in any
of the computed bands aside from a decline
in the number of brightest sources
(somewhat uncertain due to our limited box size)
and a modest increase (by about a factor of two)
in the luminosity
function for fainter objects.
The likely explanation for this has been mentioned,
it is largely due to coincidental
balancing of two effects:
new breaking waves increase the luminosity density
but mergers decrease it.
It just happens that for this spectrum at this epoch
the net rate of change is small.
In other scenarios we expect the results
to be different, but only detailed calculations can prove this.
As noted previously, our numerical results
contrast sharply with the semi-analytic, approximate computations
of Kaiser (1986) and Blanchard \etal (1992),
who were required to make many simplifying
assumptions, particularly concerning the cluster
core radii.
Kaiser (1986) assuming self-similar cluster evolution,
predicted that the luminosity
function would scale as $(1+z)^{5/2}$, in contrast
to the minimal evolution displayed in our Figures 1-4.

However, Figures 5 and 6 did show
substantial rates of change in other quantities,
the temperatures and clusters radii which are
in better accord with prior expectations.
These important trends are summarized in Figures 11 and 12,
where we see a factor of 2-3 decline
in both these quantities by redshift $1$ and a
$30\%$ decline in the (luminosity weighted) mean
temperature even in the small redshift range $z=0\rightarrow 0.2$.
An analogous velocity dispersion decline was found
by Frenk \etal (1990) in a dark matter only simulation
which would correspond to a decline
in $T_x$ with increasing redshift.
We show as dashed lines in Fig. 12 and 13 the best
fit curves
analytically predicted by Kaiser:
$R_x\propto (1+z)^{-2}$ and $T_x\propto (1+z)^{-1}$.
Examination of these figures indicates
that the radius changes less rapidly and the density more rapidly
than theoretically expected.
The trend of temperature with redshift
should be detectable even with a relatively ``soft" X-ray instrument
such as ROSAT.
It provides a test of $\Omega=1$ models
since in open, $\Omega<1$ models
the structure will tend to freeze out
at early epochs ($z\sim 1/\Omega$)
with cluster temperatures not changing
substantially after that time.

Figures (14) and (15) show the evolution in the background radiation
field in two additional ways.
The first shows what a comoving observer would
have measured at various redshifts.
Note the too large contribution to the observed
background below 1 keV.
This would have become even more extreme had we included
line emission from metals in the emissivities.
The second, Figure (15), shows the fractional contribution
to the background seen by an observer at $z=0$ in several bands
that was produced at various epochs (in integral form).
The important point to note is
that most of the X-ray background (especially
in the harder bands)
that we see locally was produced at relatively
recent ($z\le 0.5$) epochs.
This is a consequence of many things,
prominant among them the redshift factors that dilute
the observable effects of emission at high redshift.

We will return to discuss,
in greater detail,
a comparison between extant observations of cluster X-ray
sources and the theoretical simulations
in a later paper.

We are happy to acknowledge
support from grants NAGW-2448 and AST91-08103.
It is a pleasure to acknowledge NCSA for allowing
us to use their Convex-3880 supercomputer.
Discussions with Marc Davis, Patrick Henry,
Mike Norman and Simon White and an
anonymous referee are gratefully acknowledged.

\vfill\eject

\centerline{REFERENCES}
\refset
Babul, A., \& Katz, N. 1993, ApJ(Letters), 406, 251
\smallskip
\refset
Bahcall, N.A. 1993, Clusters and Groups of Galaxies, in
Astrophysical Quantities, fourth edition
\smallskip
\refset
Bardeen, J.M., Bond, J.R., Kaiser, M., and Szalay, A.S.
   1986, ApJ, 304, 15
\smallskip
\refset
Blanchard, A., Wachter, K., Evrard, A.E., \& Silk, J. 1992, ApJ, 391, 1
\smallskip
\refset
Bryan, G.L, Cen, R.Y., Norman, M. L., Ostriker, J.P, \& Stone, J. 1993, ApJ,
submitted
\smallskip
\refset
Bryan, G. L., Norman, M. L., Stone, J. M., Cen, R.Y., \& Ostriker, J.P
1994. Computer Physics Communication, in preparation.
\smallskip
\refset
Bond, J.R. 1990, in \book{The Cosmic Microwave Background: 25 years
later}, ed. N.~Mandolesi and N.~Vittorio (Dordrecht: Kluwer).
\smallskip
\refset
Bond, J.R., and Myers, S.T., 1991a, in \book{Primordial Nucleosynthesis
and Evolution of Early Universe
Physics}, ed. K. Sato and J. Audouze (Kluwer, Dordrecht), p305.
\smallskip
\refset
Bond, J.R., \& Myers, S.T. 1991b, in \book{Trends in Astroparticle
Physics}, Proceed. of the Nov. 1990 UCLA Conf., ed. D. Cline,
(Singapore: World Scientific), p24.
\smallskip
\refset
Cavaliere, A., Menci, N., \& Setti, G. 1991, Astro. Astrophy., 245, 21.
\smallskip
\refset
Cen, R.Y., Jameson, A., Liu, F., \& Ostriker, J.P.
1990, ApJ(Letters), 362, L41
\smallskip
\refset
Cen, R.Y., 1992, ApJS, 78, 341
\smallskip
\refset
Cen, R.Y., \& Ostriker, J.P, 1992, ApJ, 393, 22
\smallskip
\refset
Cen, R.Y., \& Ostriker, J.P, 1993, ApJ, in press
\smallskip
\refset
Cole, S., \& Kaiser, N. 1989, MNRAS, 237, 1127
\smallskip
\refset
Colella, P. and Woodward, P. R.  1984, J. Comp. Phys. 54, 174-201.
\smallskip
\refset
Davis, M., Efstathiou, G., Frenk, C., \& White, S.D.M. 1992, Nature, 356, 489
\smallskip
\refset
Efstathiou, G., Bond, J.R, \& White, S.D.M 1992, MNRAS, 258, 1p
\smallskip
\refset
Evrard, A.E., Summers, F., \& Davis, M. 1993, ApJ, in press
\smallskip
\refset
Evrard, A.E., \& Henry, J. P 1991, ApJ, 383, 95
\smallskip
\refset
Frenk, C.S., White, S.D.M, Efstathiou, G., \& Davis, M. 1990, ApJ, 351, 10
\smallskip
\refset
Gursky, H., Kellogg, E., Murry, S., Leony, C.,
Tananbam, H., \& Giacconi, R. 1971, ApJ(Letters), 167, L81
\smallskip
\refset
Harten, A. 1983, J. Comp. Phys., 49, 357
\smallskip
\refset
Henry, J.P. 1992, in ``Clusters and Superclusters of Galaxies",
ed. A.C. Fabian, 311, (Kluwer Academic Publisher).
\smallskip
\refset
Henry, J.P., \& Arnaud, K.A. 1991, ApJ, 372, 400
\smallskip
\refset
Jameson, A. 1989, Science, 245, 361
\smallskip
\refset
Jones, W. \& Forman, C. 1992, in ``Clusters and Superclusters of Galaxies",
ed. A.C. Fabian, 49, (Kluwer Academic Publisher).
\smallskip
\refset
Kaiser, N. 1986, MNRAS, 222, 323
\smallskip
\refset
Markevitch, M., Blumenthal, G.R., Forman, W., Jones, C.,
and Sunyaev, R.A. 1991, ApJ(Letters), 378, L33.
\smallskip
\refset
Ryu, D., Ostriker, J.P., Kang, H., \& Cen, R.Y. 1993, ApJ, 414, 1
\smallskip
\refset
Shapiro, P.R., and Struck-Marcell, P. 1985, ApJS, 57, 205
\smallskip
\refset
Smoot, G.F., \etal 1992, ApJ(Letters), 396, L1
\smallskip
\refset
Spitzer, L., Jr. 1978, Physical Processes in the Interstellar Medium
(New York: Wiley)
\smallskip
\refset
Walker, T.P., Steigman, G., Schramm, D.N., Olive, K.A., and Kang, H.S.,
   1990, 376, 51
\smallskip
\refset
White, S.D.M 1991 in ``Clusters and Superclusters of Galaxies"
(Kluwer Publishers: Dordrecht) p17, ed. A.C. Fabian
\smallskip
\refset
Wu, X., Hamilton, T., Helfand, D.J., \& Wang, Q. 1991, ApJ, 379, 564
\smallskip
\refset
Yahil, A., \& Ostriker, J.P. 1973, ApJ, 185, 787
\smallskip
\refset
Zel'dovich, Ya. 1970, Ast.Ap., 5, 84
\smallskip
\refset
Zwicky, F. 1933, Helvetica Physica Acta, 6, 110
\smallskip
\refset
\vfill\eject

\centerline{FIGURE CAPTIONS}
\medskip

\item{Fig. 1--}
Figure (1a):
the X-ray cluster bremsstrahlung luminosity function
(from central $<0.5h^{-1}Mpc$ regions)
integrated over the
whole frequency range at five different redshifts
$z=(0, 0.2, 0.5, 0.7, 1.0)$.
Figure (1b):
the X-ray cluster bremsstrahlung luminosity (from
$<1.0h^{-1}$Mpc region) function
integrated over the
whole frequency range (filled dots) at same five different redshifts.
Absence of line emission from computed luminosities
leads to an underestimate of the bolometric luminosity.
The crossed shaded area shows the observations
(Henry \& Arnaud 1991,
$\{3.1^{+4.5}_{-1.8}\times 10^{-6} h^3 {\rm Mpc}^{-3} h^2 [L_{44}({\rm
bol})]^{-1}\}
\times [h^2 L_{44}({\rm bol})]^{-1.85\pm 0.4}$).
Notice that there is minimal evolution
(for comoving observers) in contrast
to the expection (Kaiser 1986) of $(1+z)^{5/2}$ for self-similar evolution.

\item{Fig. 2--}
Same as Figure (1) but for the luminosities integrated
over $0.3-3.5$keV frequency bin.

\item{Fig. 3--}
Same as Figure (1) but for the luminosities integrated
over $0.5-4.5$keV frequency bin.
Line emission is not significant in this energy range.

\item{Fig. 4--}
Same as Figure (1) but for the luminosities integrated
over $2-10$keV frequency bin.
Solid line shows observational data taken from the review by Henry (1992).
The excess of computed over observed is significant and would
be considerably greater if we had a numerically
more accurate calculation and/or used the observed rather than
the computed gas density in clusters.

\item{Fig. 5--}
The X-ray cluster temperature ($T_x$, emission weighted temperature)
distribution function at six different redshifts $z=(0,0.5,1,2,3,5)$.
Arrows indicate the luminosity weighted average temperature $\bar T_x$
at each epoch.
Figures (a,b), as before show core and total cluster properties.
In the first ($z=0$) panel in (5b) the cross shaded area
is the observed temperature function from
Henry \& Arnaud (1991)
[$(1.8^{+0.8}_{-0.5}\times 10^{-3} h^3 {\rm Mpc}^{-3} keV^{-1})
(kT)^{-4.7\pm 0.5}$].
The artificial turnover at low $T_x$ does not effect our
luminosity weighted results.

\item{Fig. 6--}
The X-ray cluster effective radius ($r_x$)
distribution [\cf equation (2)].
Arrows indicate the luminosity weighted effective radius at each epoch.
Figures (a,b) as before.
The artificial turnover at low $r_x$ indicates
resolution problem with our
numerical methods and probably does affect our luminosity
weighted results.

\item{Fig. 7--}
Figure (7a) shows the scatter plot of $T_x$ vs $L_{tot}$ at $z=0$.
Figure (7b) shows the scatter plot of $r_x$ vs $L_{tot}$ at $z=0$;
no trend is seen
The dashed lines in upper panels show power law fits
with slopes being (0.39,0.37) for (7a,b) respectively.
The hatched region indicates the observations (Henry 1992),
$T(keV)=(0.29\pm 0.01) (h^{-2} L_{40})^{0.297\pm 0.004}$.
The difference is significant and in the expected sense.
Just as COBE normalized CDM predicts too high a galaxy
velocity dispersion,
so it predicts too high cluster X-ray temperatures.

\item{Fig. 8--}
The ratio of the central cell temperature to the temperature
of its surrounding shell ($\sim$ one cell thick) as a function
of $L_{tot}$; no trend is seen.

\item{Fig. 9--}
We compare the temperatures found
in the three regions (central cell, the shell surrounding the central
and the next outer shell)
and normalized to the temperature in the central cell.
A moderate decline of temperature with radius is found,
but there is considerable variance in the
relationship.

\item{Fig. 10--}
The ratio $\rho_{IGM}/\rho_{tot}$ as a function
of $\rho_{tot}/<\rho_{tot}>$.
Results are smoothed by a gaussian window of
radius $1h^{-1}$Mpc.
The global mean value of $\rho_{IGM}/\rho_{tot}$
is shown by the dashed line.
Note that in the highest density regions the gas is under-represented,
``anti-biased",
by a factor of about $1.7$,
whereas the voids are correspondingly gas rich.

\item{Fig. 11--}
Figure (11) shows
as the thin solid line
the distribution of
$(\rho_{gas}/\rho_{tot})$
in our tabulated clusters with the median value of this
quantity, $0.040$, indicating by the vertical thin solid line.
But, since the observed sample is (approximately) picked on a
luminosity weighted basis,
we constructed the histogram of
$(\rho_{gas}/\rho_{tot})$
using that weighting and display it as the dotted region
with median indicated by the dotted vertical thin line.
The global (assumed) average for
$(\Omega_b/\Omega_{tot})$ is shown
as the heavy vertical line.
We see an antibias of $0.67$ from the
ratio of median to global value
which is reduced to $0.75$ if
we luminosity weight the sample.
The heavy solid histogram is the observation adapted
from Jones \& Forman (1992), we see that the observed ratio
is by a factor of 2-3 higher than that computed.

\item{Fig. 12--}
The average cluster core radii in physical units as a function of redshift
for clusters with luminosity in the $0.5-4.5$keV band greater than
$10^{43}$erg/s.
The best fit evolution of the form $R_x\propto (1+z)^{-2}$ is
shown as a dotted line.

\item{Fig. 13--}
The average cluster temperature as a function of redshift
for clusters with luminosity in the $0.5-4.5$keV band greater than
$10^{43}$erg/s.
The best fit evolution of the form $T_x\propto (1+z)^{-1}$
is shown as a dotted line.

\item{Fig. 14--}
Figure (14) shows the mean radiation at six epochs,
$z=5$ (solid line),
$z=3$ (dotted line),
$z=2$ (short, dashed line),
$z=1$ (long, dashed line)
$z=0.5$ (dotted-short-dashed line) and
$z=0$ (dotted-long-dashed line).
The box in the middle shows the obseravtional data by Wu \etal (1991).
Recall that no allowance is made in our calculation for line
emission or absorption, with both important at energies
$<1$keV.
This effect, numerical effects and our too low
assumption of $\Omega_b/\Omega_{tot}$ (\cf Figure 11)
lead to a very significant underestimate of $i_\nu$
which is already in excess of what is permitted after
removal of sources.

\item{Fig. 15--}
The distribution functions of four presently observed X-ray bands
as a function of redshift (in integral form) indicating that
most observed radiation is emitted at small redshift.

\vfill\eject

\centerline {{\bf Table 1a.} Parameters of Schechter fits for the
 X-ray cluster luminosity function}
\vskip -0.3cm
\centerline {integrated over the entire frequency range.}
\vskip -0.3cm
\centerline {X-ray Cluster Core Luminosity ($<0.5h^{-1}$Mpc)}
\begintable
\hfill \quad $z$ \qquad\hfill|
\hfill \quad $\alpha$ \qquad\hfill|
\hfill \quad $L_x^*$ ($10^{44}$) \qquad\hfill|
\hfill \quad $k\bar T_x$ (keV) \qquad\hfill|
\hfill \quad $n_0$ \qquad\hfill|
\hfill \quad $n(L>10^{43})$ \qquad\hfill|
\hfill \quad $n(L>10^{44})$ \qquad\hfill|
\hfill \quad $j_{cl}$ \qquad\hfill|
\hfill \quad $j_{gas}$\qquad\hfill\cr
\hfill \quad 0 \qquad \hfill|
\hfill \quad 1.47 \qquad \hfill|
\hfill \quad 1.49 \qquad \hfill|
\hfill \quad 8.55 \qquad \hfill|
\hfill \quad 4.85 \qquad \hfill|
\hfill \quad 9.39 \qquad \hfill|
\hfill \quad 0.79 \qquad \hfill|
\hfill \quad 0.12 \qquad \hfill|
\hfill \quad 0.43 \qquad \hfill\cr
\hfill \quad 0.2 \qquad \hfill|
\hfill \quad 1.47 \qquad \hfill|
\hfill \quad 1.02 \qquad \hfill|
\hfill \quad 4.88 \qquad \hfill|
\hfill \quad 9.36 \qquad \hfill|
\hfill \quad 13.5 \qquad \hfill|
\hfill \quad 0.76 \qquad \hfill|
\hfill \quad 0.16 \qquad \hfill|
\hfill \quad 0.52 \qquad \hfill\cr
\hfill \quad 0.5 \qquad \hfill|
\hfill \quad 1.47 \qquad \hfill|
\hfill \quad 0.52 \qquad \hfill|
\hfill \quad 5.20 \qquad \hfill|
\hfill \quad 20.1 \qquad \hfill|
\hfill \quad 10.8 \qquad \hfill|
\hfill \quad 0.39 \qquad \hfill|
\hfill \quad 0.18 \qquad \hfill|
\hfill \quad 0.62 \qquad \hfill\cr
\hfill \quad 0.7 \qquad \hfill|
\hfill \quad 1.51 \qquad \hfill|
\hfill \quad 0.68 \qquad \hfill|
\hfill \quad 3.14 \qquad \hfill|
\hfill \quad 18.0 \qquad \hfill|
\hfill \quad 19.1 \qquad \hfill|
\hfill \quad 0.57 \qquad \hfill|
\hfill \quad 0.22 \qquad \hfill|
\hfill \quad 0.62 \qquad \hfill\cr
\hfill \quad 1 \qquad \hfill|
\hfill \quad 1.48 \qquad \hfill|
\hfill \quad 0.46 \qquad \hfill|
\hfill \quad 2.90 \qquad \hfill|
\hfill \quad 30.1 \qquad \hfill|
\hfill \quad 21.2 \qquad \hfill|
\hfill \quad 0.31 \qquad \hfill|
\hfill \quad 0.24 \qquad \hfill|
\hfill \quad 0.62 \qquad \hfill\cr
\hfill \quad 2 \qquad \hfill|
\hfill \quad 1.59 \qquad \hfill|
\hfill \quad 0.46 \qquad \hfill|
\hfill \quad 1.71 \qquad \hfill|
\hfill \quad 19.2 \qquad \hfill|
\hfill \quad 14.8 \qquad \hfill|
\hfill \quad 0.17 \qquad \hfill|
\hfill \quad 0.19 \qquad \hfill|
\hfill \quad 0.48 \qquad \hfill\cr
\hfill \quad 3 \qquad \hfill|
\hfill \quad 1.63 \qquad \hfill|
\hfill \quad 0.24 \qquad \hfill|
\hfill \quad 0.71 \qquad \hfill|
\hfill \quad 34.1 \qquad \hfill|
\hfill \quad 11.8 \qquad \hfill|
\hfill \quad 0.017 \qquad \hfill|
\hfill \quad 0.20 \qquad \hfill|
\hfill \quad 0.45 \qquad \hfill\cr
\hfill \quad 5 \qquad \hfill|
\hfill \quad 1.97 \qquad \hfill|
\hfill \quad 0.16 \qquad \hfill|
\hfill \quad 0.20 \qquad \hfill|
\hfill \quad 5.39 \qquad \hfill|
\hfill \quad 0.99 \qquad \hfill|
\hfill \quad $9.6\times 10^{-5}$ \qquad \hfill|
\hfill \quad 0.28 \qquad \hfill|
\hfill \quad 0.17 \qquad \hfill
\endtable
\centerline {X-ray Cluster Total Luminosity ($<1h^{-1}$Mpc)}
\begintable
\hfill \quad $z$ \qquad\hfill|
\hfill \quad $\alpha$ \qquad\hfill|
\hfill \quad $L_x^*$ ($10^{44}$) \qquad\hfill|
\hfill \quad $k\bar T_x$ (keV) \qquad\hfill|
\hfill \quad $n_0$ \qquad\hfill|
\hfill \quad $n(L>10^{43})$ \qquad\hfill|
\hfill \quad $n(L>10^{44})$ \qquad\hfill|
\hfill \quad $j_{cl}$ \qquad\hfill|
\hfill \quad $j_{gas}$\qquad\hfill\cr
\hfill \quad 0 \qquad \hfill|
\hfill \quad 1.44 \qquad \hfill|
\hfill \quad 3.33 \qquad \hfill|
\hfill \quad 6.08 \qquad \hfill|
\hfill \quad 5.23 \qquad \hfill|
\hfill \quad 16.5 \qquad \hfill|
\hfill \quad 2.53 \qquad \hfill|
\hfill \quad 0.28 \qquad \hfill|
\hfill \quad 0.43 \qquad \hfill\cr
\hfill \quad 0.2 \qquad \hfill|
\hfill \quad 1.43 \qquad \hfill|
\hfill \quad 1.89 \qquad \hfill|
\hfill \quad 3.98 \qquad \hfill|
\hfill \quad 12.1 \qquad \hfill|
\hfill \quad 25.9 \qquad \hfill|
\hfill \quad 2.84 \qquad \hfill|
\hfill \quad 0.36 \qquad \hfill|
\hfill \quad 0.52 \qquad \hfill\cr
\hfill \quad 0.5 \qquad \hfill|
\hfill \quad 1.44 \qquad \hfill|
\hfill \quad 1.30 \qquad \hfill|
\hfill \quad 3.56 \qquad \hfill|
\hfill \quad 18.9 \qquad \hfill|
\hfill \quad 31.5 \qquad \hfill|
\hfill \quad 2.45 \qquad \hfill|
\hfill \quad 0.39 \qquad \hfill|
\hfill \quad 0.62 \qquad \hfill\cr
\hfill \quad 0.7 \qquad \hfill|
\hfill \quad 1.48 \qquad \hfill|
\hfill \quad 1.88 \qquad \hfill|
\hfill \quad 2.51 \qquad \hfill|
\hfill \quad 14.8 \qquad \hfill|
\hfill \quad 34.6 \qquad \hfill|
\hfill \quad 3.46 \qquad \hfill|
\hfill \quad 0.47 \qquad \hfill|
\hfill \quad 0.62 \qquad \hfill\cr
\hfill \quad 1 \qquad \hfill|
\hfill \quad 1.44 \qquad \hfill|
\hfill \quad 0.93 \qquad \hfill|
\hfill \quad 1.98 \qquad \hfill|
\hfill \quad 33.0 \qquad \hfill|
\hfill \quad 42.5 \qquad \hfill|
\hfill \quad 2.26 \qquad \hfill|
\hfill \quad 0.49 \qquad \hfill|
\hfill \quad 0.62 \qquad \hfill\cr
\hfill \quad 2 \qquad \hfill|
\hfill \quad 1.53 \qquad \hfill|
\hfill \quad 0.56 \qquad \hfill|
\hfill \quad 1.15 \qquad \hfill|
\hfill \quad 32.5 \qquad \hfill|
\hfill \quad 29.2 \qquad \hfill|
\hfill \quad 0.59 \qquad \hfill|
\hfill \quad 0.34 \qquad \hfill|
\hfill \quad 0.48 \qquad \hfill\cr
\hfill \quad 3 \qquad \hfill|
\hfill \quad 1.53 \qquad \hfill|
\hfill \quad 0.26 \qquad \hfill|
\hfill \quad 0.54 \qquad \hfill|
\hfill \quad 67.5 \qquad \hfill|
\hfill \quad 25.1 \qquad \hfill|
\hfill \quad 0.060 \qquad \hfill|
\hfill \quad 0.33 \qquad \hfill|
\hfill \quad 0.45 \qquad \hfill\cr
\hfill \quad 5 \qquad \hfill|
\hfill \quad 1.91 \qquad \hfill|
\hfill \quad 0.20 \qquad \hfill|
\hfill \quad 0.17 \qquad \hfill|
\hfill \quad 9.62 \qquad \hfill|
\hfill \quad 2.68 \qquad \hfill|
\hfill \quad $9.8\times 10^{-4}$ \qquad \hfill|
\hfill \quad 0.20 \qquad \hfill|
\hfill \quad 0.17 \qquad \hfill
\endtable
\vskip -0.4cm
\ninerm Here
$L_x^*$ and $L$ are in units of $10^{44}$erg/s;
$n_0$ and $n(L>10^{44})$ are in units of $10^{-6}h^3$Mpc$^{-3}$;
$j_{cl}$ and $j_{gas}$ are in
units of $10^{40}$erg/s/h$^{-3}$Mpc$^3$, and $j_{cl}$ may be larger
than $j_{gas}$ due to the inaccuracy of the Schechter fit.
{}.
\vfill\eject

\centerline {{\bf Table 1b.} Parameters of Schechter fits for the
 X-ray cluster luminosity function}
\vskip -0.3cm
\centerline {in $0.3-3.5$keV band}
\vskip -0.3cm
\centerline {X-ray Cluster Core Luminosity ($<0.5h^{-1}$Mpc)}
\begintable
\hfill \quad $z$ \qquad\hfill|
\hfill \quad $\alpha$ \qquad\hfill|
\hfill \quad $L_x^*$ \qquad\hfill|
\hfill \quad $n_0$ \qquad\hfill|
\hfill \quad $n(L>10^{43})$ \qquad\hfill|
\hfill \quad $n(L>10^{44})$ \qquad\hfill|
\hfill \quad $j_{cl}$ \qquad\hfill|
\hfill \quad $j_{gas}$ \qquad\hfill\cr
\hfill \quad 0 \qquad \hfill|
\hfill \quad 1.41 \qquad \hfill|
\hfill \quad 0.25 \qquad \hfill|
\hfill \quad 11.9 \qquad \hfill|
\hfill \quad 4.04 \qquad \hfill|
\hfill \quad $1.0\times 10^{-2}$ \qquad \hfill|
\hfill \quad 0.045 \qquad \hfill|
\hfill \quad 0.16 \qquad \hfill\cr
\hfill \quad 0.2 \qquad \hfill|
\hfill \quad 1.39 \qquad \hfill|
\hfill \quad 0.27 \qquad \hfill|
\hfill \quad 20.0 \qquad \hfill|
\hfill \quad 8.40 \qquad \hfill|
\hfill \quad $6.0\times 10^{-2}$ \qquad \hfill|
\hfill \quad 0.079 \qquad \hfill|
\hfill \quad 0.24 \qquad \hfill\cr
\hfill \quad 0.5 \qquad \hfill|
\hfill \quad 1.43 \qquad \hfill|
\hfill \quad 0.28 \qquad \hfill|
\hfill \quad 22.6 \qquad \hfill|
\hfill \quad 9.91 \qquad \hfill|
\hfill \quad $2.8\times 10^{-2}$ \qquad \hfill|
\hfill \quad 0.099 \qquad \hfill|
\hfill \quad 0.30 \qquad \hfill\cr
\hfill \quad 0.7 \qquad \hfill|
\hfill \quad 1.34 \qquad \hfill|
\hfill \quad 0.15 \qquad \hfill|
\hfill \quad 56.1 \qquad \hfill|
\hfill \quad 13.2 \qquad \hfill|
\hfill \quad $8.0\times 10^{-2}$ \qquad \hfill|
\hfill \quad 0.11 \qquad \hfill|
\hfill \quad 0.33 \qquad \hfill\cr
\hfill \quad 1 \qquad \hfill|
\hfill \quad 1.34 \qquad \hfill|
\hfill \quad 0.14 \qquad \hfill|
\hfill \quad 65.3 \qquad \hfill|
\hfill \quad 13.3 \qquad \hfill|
\hfill \quad $1.6\times 10^{-2}$ \qquad \hfill|
\hfill \quad 0.12 \qquad \hfill|
\hfill \quad 0.34 \qquad \hfill\cr
\hfill \quad 2 \qquad \hfill|
\hfill \quad 1.41 \qquad \hfill|
\hfill \quad 0.15 \qquad \hfill|
\hfill \quad 41.0 \qquad \hfill|
\hfill \quad 6.73 \qquad \hfill|
\hfill \quad $4.5\times 10^{-2}$ \qquad \hfill|
\hfill \quad 0.093 \qquad \hfill|
\hfill \quad 0.25 \qquad \hfill\cr
\hfill \quad 3 \qquad \hfill|
\hfill \quad 1.31 \qquad \hfill|
\hfill \quad 0.089 \qquad \hfill|
\hfill \quad 78.3 \qquad \hfill|
\hfill \quad 2.53 \qquad \hfill|
\hfill \quad $2.2\times 10^{-4}$ \qquad \hfill|
\hfill \quad 0.091 \qquad \hfill|
\hfill \quad 0.19 \qquad \hfill\cr
\hfill \quad 5 \qquad \hfill|
\hfill \quad 1.65 \qquad \hfill|
\hfill \quad 0.14 \qquad \hfill|
\hfill \quad 2.92 \qquad \hfill|
\hfill \quad $3.0\times 10^{-5}$ \qquad \hfill|
\hfill \quad $1.0\times 10^{-7}$ \qquad \hfill|
\hfill \quad 0.010 \qquad \hfill|
\hfill \quad 0.019 \qquad \hfill
\endtable
\centerline {X-ray Cluster Total Luminosity ($<1h^{-1}$Mpc)}
\begintable
\hfill \quad $z$ \qquad\hfill|
\hfill \quad $\alpha$ \qquad\hfill|
\hfill \quad $L_x^*$ \qquad\hfill|
\hfill \quad $n_0$ \qquad\hfill|
\hfill \quad $n(L>10^{43})$ \qquad\hfill|
\hfill \quad $n(L>10^{44})$ \qquad\hfill|
\hfill \quad $j_{cl}$ \qquad\hfill|
\hfill \quad $j_{gas}$ \qquad\hfill\cr
\hfill \quad 0 \qquad \hfill|
\hfill \quad 1.36 \qquad \hfill|
\hfill \quad 0.64 \qquad \hfill|
\hfill \quad 13.2 \qquad \hfill|
\hfill \quad 11.4 \qquad \hfill|
\hfill \quad $0.40$ \qquad \hfill|
\hfill \quad 0.12 \qquad \hfill|
\hfill \quad 0.16 \qquad \hfill\cr
\hfill \quad 0.2 \qquad \hfill|
\hfill \quad 1.33 \qquad \hfill|
\hfill \quad 0.60 \qquad \hfill|
\hfill \quad 23.3 \qquad \hfill|
\hfill \quad 18.5 \qquad \hfill|
\hfill \quad $0.61$ \qquad \hfill|
\hfill \quad 0.19 \qquad \hfill|
\hfill \quad 0.24 \qquad \hfill\cr
\hfill \quad 0.5 \qquad \hfill|
\hfill \quad 1.36 \qquad \hfill|
\hfill \quad 0.53 \qquad \hfill|
\hfill \quad 28.8 \qquad \hfill|
\hfill \quad 21.1 \qquad \hfill|
\hfill \quad $0.51$ \qquad \hfill|
\hfill \quad 0.21 \qquad \hfill|
\hfill \quad 0.30 \qquad \hfill\cr
\hfill \quad 0.7 \qquad \hfill|
\hfill \quad 1.29 \qquad \hfill|
\hfill \quad 0.32 \qquad \hfill|
\hfill \quad 58.7 \qquad \hfill|
\hfill \quad 25.2 \qquad \hfill|
\hfill \quad $0.19$ \qquad \hfill|
\hfill \quad 0.24 \qquad \hfill|
\hfill \quad 0.33 \qquad \hfill\cr
\hfill \quad 1 \qquad \hfill|
\hfill \quad 1.29 \qquad \hfill|
\hfill \quad 0.30 \qquad \hfill|
\hfill \quad 66.8 \qquad \hfill|
\hfill \quad 26.8 \qquad \hfill|
\hfill \quad $0.17$ \qquad \hfill|
\hfill \quad 0.26 \qquad \hfill|
\hfill \quad 0.34 \qquad \hfill\cr
\hfill \quad 2 \qquad \hfill|
\hfill \quad 1.31 \qquad \hfill|
\hfill \quad 0.24 \qquad \hfill|
\hfill \quad 59.2 \qquad \hfill|
\hfill \quad 18.6 \qquad \hfill|
\hfill \quad $0.049$ \qquad \hfill|
\hfill \quad 0.19 \qquad \hfill|
\hfill \quad 0.25 \qquad \hfill\cr
\hfill \quad 3 \qquad \hfill|
\hfill \quad 1.33 \qquad \hfill|
\hfill \quad 0.20 \qquad \hfill|
\hfill \quad 53.1 \qquad \hfill|
\hfill \quad 13.3 \qquad \hfill|
\hfill \quad $0.015$ \qquad \hfill|
\hfill \quad 0.14 \qquad \hfill|
\hfill \quad 0.19 \qquad \hfill\cr
\hfill \quad 5 \qquad \hfill|
\hfill \quad 1.58 \qquad \hfill|
\hfill \quad 0.23 \qquad \hfill|
\hfill \quad 3.56 \qquad \hfill|
\hfill \quad 1.14 \qquad \hfill|
\hfill \quad $1.5\times 10^{-3}$ \qquad \hfill|
\hfill \quad 0.017 \qquad \hfill|
\hfill \quad 0.019 \qquad \hfill
\endtable
\vfill\eject

\centerline {{\bf Table 1c.} Parameters of Schechter fits for the
 X-ray cluster luminosity function}
\vskip -0.3cm
\centerline {in $0.5-4.5$keV band}
\vskip -0.3cm
\centerline {X-ray Cluster Core Luminosity ($<0.5h^{-1}$Mpc)}
\begintable
\hfill \quad $z$ \qquad\hfill|
\hfill \quad $\alpha$ \qquad\hfill|
\hfill \quad $L_x^*$ \qquad\hfill|
\hfill \quad $n_0$ \qquad\hfill|
\hfill \quad $n(L>10^{43})$ \qquad\hfill|
\hfill \quad $n(L>10^{44})$ \qquad\hfill|
\hfill \quad $j_{cl}$ \qquad\hfill|
\hfill \quad $j_{gas}$ \qquad\hfill\cr
\hfill \quad 0 \qquad \hfill|
\hfill \quad 1.36 \qquad \hfill|
\hfill \quad 0.23 \qquad \hfill|
\hfill \quad 14.1 \qquad \hfill|
\hfill \quad 4.26 \qquad \hfill|
\hfill \quad $8.5\times 10^{-3}$ \qquad \hfill|
\hfill \quad 0.046 \qquad \hfill|
\hfill \quad 0.18 \qquad \hfill\cr
\hfill \quad 0.2 \qquad \hfill|
\hfill \quad 1.34 \qquad \hfill|
\hfill \quad 0.27 \qquad \hfill|
\hfill \quad 21.0 \qquad \hfill|
\hfill \quad 8.61 \qquad \hfill|
\hfill \quad $7.9\times 10^{-2}$ \qquad \hfill|
\hfill \quad 0.077 \qquad \hfill|
\hfill \quad 0.24 \qquad \hfill\cr
\hfill \quad 0.5 \qquad \hfill|
\hfill \quad 1.37 \qquad \hfill|
\hfill \quad 0.26 \qquad \hfill|
\hfill \quad 25.4 \qquad \hfill|
\hfill \quad 8.94 \qquad \hfill|
\hfill \quad $2.9\times 10^{-2}$ \qquad \hfill|
\hfill \quad 0.094 \qquad \hfill|
\hfill \quad 0.29 \qquad \hfill\cr
\hfill \quad 0.7 \qquad \hfill|
\hfill \quad 1.31 \qquad \hfill|
\hfill \quad 0.16 \qquad \hfill|
\hfill \quad 52.8 \qquad \hfill|
\hfill \quad 12.0 \qquad \hfill|
\hfill \quad $9.2\times 10^{-2}$ \qquad \hfill|
\hfill \quad 0.11 \qquad \hfill|
\hfill \quad 0.31 \qquad \hfill\cr
\hfill \quad 1 \qquad \hfill|
\hfill \quad 1.30 \qquad \hfill|
\hfill \quad 0.15 \qquad \hfill|
\hfill \quad 61.8 \qquad \hfill|
\hfill \quad 10.4 \qquad \hfill|
\hfill \quad $1.0\times 10^{-2}$ \qquad \hfill|
\hfill \quad 0.12 \qquad \hfill|
\hfill \quad 0.31 \qquad \hfill\cr
\hfill \quad 2 \qquad \hfill|
\hfill \quad 1.39 \qquad \hfill|
\hfill \quad 0.17 \qquad \hfill|
\hfill \quad 33.3 \qquad \hfill|
\hfill \quad 5.43 \qquad \hfill|
\hfill \quad $4.9\times 10^{-2}$ \qquad \hfill|
\hfill \quad 0.083 \qquad \hfill|
\hfill \quad 0.20 \qquad \hfill\cr
\hfill \quad 3 \qquad \hfill|
\hfill \quad 1.31 \qquad \hfill|
\hfill \quad 0.081 \qquad \hfill|
\hfill \quad 60.1 \qquad \hfill|
\hfill \quad 0.89 \qquad \hfill|
\hfill \quad $1.0\times 10^{-5}$ \qquad \hfill|
\hfill \quad 0.064 \qquad \hfill|
\hfill \quad 0.13 \qquad \hfill\cr
\hfill \quad 5 \qquad \hfill|
\hfill \quad 1.65 \qquad \hfill|
\hfill \quad 0.10 \qquad \hfill|
\hfill \quad 1.6 \qquad \hfill|
\hfill \quad 0.11 \qquad \hfill|
\hfill \quad $1.0\times 10^{-7}$ \qquad \hfill|
\hfill \quad 0.004 \qquad \hfill|
\hfill \quad 0.010 \qquad \hfill
\endtable
\centerline {X-ray Cluster Total Luminosity ($<1h^{-1}$Mpc)}
\begintable
\hfill \quad $z$ \qquad\hfill|
\hfill \quad $\alpha$ \qquad\hfill|
\hfill \quad $L_x^*$ \qquad\hfill|
\hfill \quad $n_0$ \qquad\hfill|
\hfill \quad $n(L>10^{43})$ \qquad\hfill|
\hfill \quad $n(L>10^{44})$ \qquad\hfill|
\hfill \quad $j_{cl}$ \qquad\hfill|
\hfill \quad $j_{gas}$ \qquad\hfill\cr
\hfill \quad 0 \qquad \hfill|
\hfill \quad 1.34 \qquad \hfill|
\hfill \quad 0.68 \qquad \hfill|
\hfill \quad 12.7 \qquad \hfill|
\hfill \quad 11.3 \qquad \hfill|
\hfill \quad $0.45$ \qquad \hfill|
\hfill \quad 0.12 \qquad \hfill|
\hfill \quad 0.18 \qquad \hfill\cr
\hfill \quad 0.2 \qquad \hfill|
\hfill \quad 1.32 \qquad \hfill|
\hfill \quad 0.82 \qquad \hfill|
\hfill \quad 19.0 \qquad \hfill|
\hfill \quad 19.1 \qquad \hfill|
\hfill \quad $1.06$ \qquad \hfill|
\hfill \quad 0.21 \qquad \hfill|
\hfill \quad 0.24 \qquad \hfill\cr
\hfill \quad 0.5 \qquad \hfill|
\hfill \quad 1.32 \qquad \hfill|
\hfill \quad 0.54 \qquad \hfill|
\hfill \quad 29.2 \qquad \hfill|
\hfill \quad 21.0 \qquad \hfill|
\hfill \quad $0.57$ \qquad \hfill|
\hfill \quad 0.21 \qquad \hfill|
\hfill \quad 0.29 \qquad \hfill\cr
\hfill \quad 0.7 \qquad \hfill|
\hfill \quad 1.26 \qquad \hfill|
\hfill \quad 0.33 \qquad \hfill|
\hfill \quad 55.8 \qquad \hfill|
\hfill \quad 24.4 \qquad \hfill|
\hfill \quad $0.22$ \qquad \hfill|
\hfill \quad 0.23 \qquad \hfill|
\hfill \quad 0.31 \qquad \hfill\cr
\hfill \quad 1 \qquad \hfill|
\hfill \quad 1.26 \qquad \hfill|
\hfill \quad 0.31 \qquad \hfill|
\hfill \quad 61.7 \qquad \hfill|
\hfill \quad 25.3 \qquad \hfill|
\hfill \quad $0.18$ \qquad \hfill|
\hfill \quad 0.24 \qquad \hfill|
\hfill \quad 0.31 \qquad \hfill\cr
\hfill \quad 2 \qquad \hfill|
\hfill \quad 1.32 \qquad \hfill|
\hfill \quad 0.31 \qquad \hfill|
\hfill \quad 38.1 \qquad \hfill|
\hfill \quad 16.1 \qquad \hfill|
\hfill \quad $0.11$ \qquad \hfill|
\hfill \quad 0.16 \qquad \hfill|
\hfill \quad 0.20 \qquad \hfill\cr
\hfill \quad 3 \qquad \hfill|
\hfill \quad 1.32 \qquad \hfill|
\hfill \quad 0.22 \qquad \hfill|
\hfill \quad 37.4 \qquad \hfill|
\hfill \quad 10.6 \qquad \hfill|
\hfill \quad $0.019$ \qquad \hfill|
\hfill \quad 0.11 \qquad \hfill|
\hfill \quad 0.13 \qquad \hfill\cr
\hfill \quad 5 \qquad \hfill|
\hfill \quad 1.58 \qquad \hfill|
\hfill \quad 0.34 \qquad \hfill|
\hfill \quad 1.35 \qquad \hfill|
\hfill \quad 0.72 \qquad \hfill|
\hfill \quad $3.9\times 10^{-3}$ \qquad \hfill|
\hfill \quad 0.0097 \qquad \hfill|
\hfill \quad 0.010 \qquad \hfill
\endtable
\vfill\eject

\centerline {{\bf Table 1d.} Parameters of Schechter fits for the
 X-ray cluster luminosity function}
\vskip -0.3cm
\centerline {in $2-10$keV band}
\vskip -0.3cm
\centerline {X-ray Cluster Core Luminosity ($<0.5h^{-1}$Mpc)}
\begintable
\hfill \quad $z$ \qquad\hfill|
\hfill \quad $\alpha$ \qquad\hfill|
\hfill \quad $L_x^*$ \qquad\hfill|
\hfill \quad $n_0$ \qquad\hfill|
\hfill \quad $n(L>10^{43})$ \qquad\hfill|
\hfill \quad $n(L>10^{44})$ \qquad\hfill|
\hfill \quad $j_{cl}$ \qquad\hfill|
\hfill \quad $j_{gas}$ \qquad\hfill\cr
\hfill \quad 0 \qquad \hfill|
\hfill \quad 1.33 \qquad \hfill|
\hfill \quad 0.87 \qquad \hfill|
\hfill \quad 6.26 \qquad \hfill|
\hfill \quad 6.65 \qquad \hfill|
\hfill \quad $0.39$ \qquad \hfill|
\hfill \quad 0.073 \qquad \hfill|
\hfill \quad 0.18 \qquad \hfill\cr
\hfill \quad 0.2 \qquad \hfill|
\hfill \quad 1.27 \qquad \hfill|
\hfill \quad 0.35 \qquad \hfill|
\hfill \quad 15.4 \qquad \hfill|
\hfill \quad 12.0 \qquad \hfill|
\hfill \quad $0.50$ \qquad \hfill|
\hfill \quad 0.068 \qquad \hfill|
\hfill \quad 0.21 \qquad \hfill\cr
\hfill \quad 0.5 \qquad \hfill|
\hfill \quad 1.28 \qquad \hfill|
\hfill \quad 0.30 \qquad \hfill|
\hfill \quad 18.8 \qquad \hfill|
\hfill \quad 13.5 \qquad \hfill|
\hfill \quad $0.56$ \qquad \hfill|
\hfill \quad 0.071 \qquad \hfill|
\hfill \quad 0.29 \qquad \hfill\cr
\hfill \quad 0.7 \qquad \hfill|
\hfill \quad 1.29 \qquad \hfill|
\hfill \quad 0.29 \qquad \hfill|
\hfill \quad 21.3 \qquad \hfill|
\hfill \quad 16.6 \qquad \hfill|
\hfill \quad $0.35$ \qquad \hfill|
\hfill \quad 0.079 \qquad \hfill|
\hfill \quad 0.21 \qquad \hfill\cr
\hfill \quad 1 \qquad \hfill|
\hfill \quad 1.28 \qquad \hfill|
\hfill \quad 0.24 \qquad \hfill|
\hfill \quad 25.9 \qquad \hfill|
\hfill \quad 17.6 \qquad \hfill|
\hfill \quad $0.32$ \qquad \hfill|
\hfill \quad 0.079 \qquad \hfill|
\hfill \quad 0.18 \qquad \hfill\cr
\hfill \quad 2 \qquad \hfill|
\hfill \quad 1.46 \qquad \hfill|
\hfill \quad 0.70 \qquad \hfill|
\hfill \quad 3.69 \qquad \hfill|
\hfill \quad 7.31 \qquad \hfill|
\hfill \quad $0.057$ \qquad \hfill|
\hfill \quad 0.042 \qquad \hfill|
\hfill \quad 0.080 \qquad \hfill\cr
\hfill \quad 3 \qquad \hfill|
\hfill \quad 1.30 \qquad \hfill|
\hfill \quad 0.05 \qquad \hfill|
\hfill \quad 20.0 \qquad \hfill|
\hfill \quad 2.43 \qquad \hfill|
\hfill \quad 0.020 \qquad \hfill|
\hfill \quad 0.013 \qquad \hfill|
\hfill \quad 0.026 \qquad \hfill\cr
\hfill \quad 5 \qquad \hfill|
\hfill \quad 1.47 \qquad \hfill|
\hfill \quad 0.018 \qquad \hfill|
\hfill \quad 0.51 \qquad \hfill|
\hfill \quad $1.8\times 10^{-3}$ \qquad \hfill|
\hfill \quad $1.0\times 10^{-7}$ \qquad \hfill|
\hfill \quad $1.5\times 10^{-4}$ \qquad \hfill|
\hfill \quad $7.6\times 10^{-4}$ \qquad \hfill
\endtable
\centerline {X-ray Cluster Total Luminosity ($<1h^{-1}$Mpc)}
\begintable
\hfill \quad $z$ \qquad\hfill|
\hfill \quad $\alpha$ \qquad\hfill|
\hfill \quad $L_x^*$ \qquad\hfill|
\hfill \quad $n_0$ \qquad\hfill|
\hfill \quad $n(L>10^{43})$ \qquad\hfill|
\hfill \quad $n(L>10^{44})$ \qquad\hfill|
\hfill \quad $j_{cl}$ \qquad\hfill|
\hfill \quad $j_{gas}$ \qquad\hfill\cr
\hfill \quad 0 \qquad \hfill|
\hfill \quad 1.28 \qquad \hfill|
\hfill \quad 2.84 \qquad \hfill|
\hfill \quad 6.45 \qquad \hfill|
\hfill \quad 13.2 \qquad \hfill|
\hfill \quad $2.43$ \qquad \hfill|
\hfill \quad 0.23 \qquad \hfill|
\hfill \quad 0.18 \qquad \hfill\cr
\hfill \quad 0.2 \qquad \hfill|
\hfill \quad 1.23 \qquad \hfill|
\hfill \quad 0.72 \qquad \hfill|
\hfill \quad 17.0 \qquad \hfill|
\hfill \quad 14.1 \qquad \hfill|
\hfill \quad $0.75$ \qquad \hfill|
\hfill \quad 0.15 \qquad \hfill|
\hfill \quad 0.21 \qquad \hfill\cr
\hfill \quad 0.5 \qquad \hfill|
\hfill \quad 1.25 \qquad \hfill|
\hfill \quad 0.64 \qquad \hfill|
\hfill \quad 19.0 \qquad \hfill|
\hfill \quad 14.7 \qquad \hfill|
\hfill \quad $0.62$ \qquad \hfill|
\hfill \quad 0.15 \qquad \hfill|
\hfill \quad 0.29 \qquad \hfill\cr
\hfill \quad 0.7 \qquad \hfill|
\hfill \quad 1.28 \qquad \hfill|
\hfill \quad 0.80 \qquad \hfill|
\hfill \quad 16.5 \qquad \hfill|
\hfill \quad 15.6 \qquad \hfill|
\hfill \quad $0.89$ \qquad \hfill|
\hfill \quad 0.17 \qquad \hfill|
\hfill \quad 0.21 \qquad \hfill\cr
\hfill \quad 1 \qquad \hfill|
\hfill \quad 1.26 \qquad \hfill|
\hfill \quad 0.53 \qquad \hfill|
\hfill \quad 24.2 \qquad \hfill|
\hfill \quad 16.3 \qquad \hfill|
\hfill \quad $0.47$ \qquad \hfill|
\hfill \quad 0.16 \qquad \hfill|
\hfill \quad 0.18 \qquad \hfill\cr
\hfill \quad 2 \qquad \hfill|
\hfill \quad 1.34 \qquad \hfill|
\hfill \quad 0.38 \qquad \hfill|
\hfill \quad 11.0 \qquad \hfill|
\hfill \quad 5.78 \qquad \hfill|
\hfill \quad $0.067$ \qquad \hfill|
\hfill \quad 0.057 \qquad \hfill|
\hfill \quad 0.080 \qquad \hfill\cr
\hfill \quad 3 \qquad \hfill|
\hfill \quad 1.26 \qquad \hfill|
\hfill \quad 0.10 \qquad \hfill|
\hfill \quad 16.0 \qquad \hfill|
\hfill \quad 1.37 \qquad \hfill|
\hfill \quad $1.6\times 10^{-5}$ \qquad \hfill|
\hfill \quad 0.020 \qquad \hfill|
\hfill \quad 0.026 \qquad \hfill\cr
\hfill \quad 5 \qquad \hfill|
\hfill \quad 1.27 \qquad \hfill|
\hfill \quad 0.020 \qquad \hfill|
\hfill \quad 1.80 \qquad \hfill|
\hfill \quad $5.6\times 10^{-4}$ \qquad \hfill|
\hfill \quad $1.9\times 10^{-22}$ \qquad \hfill|
\hfill \quad $4.5\times 10^{-4}$ \qquad \hfill|
\hfill \quad $7.6\times 10^{-4}$ \qquad \hfill
\endtable

\vfill\eject\end